\documentclass[11pt]{article}
\usepackage{amsmath}
\usepackage{amssymb}
\usepackage{ntheorem}
\usepackage{graphicx}
\usepackage{subfigure}
\usepackage{color}
\usepackage{bm}
\usepackage{comment}

\numberwithin{equation}{section}
\newtheorem{theorem}{Theorem}[section]
\newtheorem{proposition}[theorem]{Proposition}
\newtheorem{definition}[theorem]{Definition}
\newtheorem{lemma}[theorem]{Lemma}
\newtheorem{remark}[theorem]{Remark}
\def\Proof{\noindent{\bf Proof} \quad}
\def\qed{\hfill $\Box$ \smallskip}

\DeclareMathOperator{\Span}{Span}
\DeclareMathOperator{\Ker}{Ker}
\DeclareMathOperator{\Det}{Det}
\DeclareMathOperator{\diag}{diag}
\DeclareMathOperator{\sgn}{sgn}

\def\N{{\mathbb N}}
\def\Z{{\mathbb Z}}

\def\R{{\mathbb R}}

\def\bk{{\bf k}}

\def\mcd{{\mathcal D}}
\def\mce{{\mathcal E}}

\def\mch{{\mathcal H}}
\def\mcl{{\mathcal L}}

\def\mcn{{\mathcal N}}
\def\mcr{{\mathcal R}}

\def\mct{{\mathcal T}}

\def\mcp{{\mathcal P}}
\def\mcf{{\mathcal F}}
\def\mcq{{\mathcal Q}}
\def\mcv{{\mathcal V}}
\def\mcs{{\mathcal S}}

\def\bx{{\bf x}}
\def\bx{{\bf x}}
\def\by{{\bf y}}
\def\bv{{\bf v}}

\def\bq{{\bf q}}
\def\br{{\bf r}}
\def\buu{{\bf u}}

\def\q{{\quad}}
\def\qq{{\qquad}}

\begin{document}

\title{Edge states in super honeycomb  structures with $\mcp\mct$ symmetric deformations}

\author{Ying Cao\thanks{Yau Mathematical Sciences Center, Tsinghua Unversity, Beijing, 100084, China ({caoy20@mails.tsinghua.edu.cn}).}, \and Yi Zhu \thanks{Yau Mathematical Sciences Center, Tsinghua Unversity, Beijing, 100084, China, and Beijing Institute of Mathematical Sciences and Applications, Beijing, 101408, China({yizhu@tsinghua.edu.cn}).} }

\maketitle

\begin{abstract}
The existence of edge states is one of the most vital properties of topological insulators. Although tremendous success has been accomplished in describing and explaining edge states associated with $\mcp\mct$ symmetry breaking, little work has been done on $\mcp\mct$ symmetry preserving cases. Two-dimensional Schr\"odinger operators with super honeycomb lattice potentials always have double Dirac cones at the $\Gamma$ point -- the zero momentum point on their energy bands due to { a rotation symmetry}, $\mcp\mct$ symmetry, and the ``folding" symmetry -- caused by an additional translation symmetry. There are two topologically different ways to deform such a system by $\mcp\mct$ symmetry preserving but folding symmetry breaking perturbations. Interestingly, there exist two gapped edge states on the interface between such two kinds of perturbed materials. In this paper, we illustrate the existence of such $\mcp\mct$ preserving edge states rigorously for the first time. We use a domain wall modulated Schr\"odinger operator to model the phenomenon under small perturbations and rigorously prove the existence of two gapped edge states. {We also provide topological interpretations from the point of view of local topological charges and the parities of degenerate bulk modes.} Our work thoroughly explains the existence of ``helical" like edge states in super honeycomb configurations and lays a foundation for the descriptions of topologies of such systems.  
\end{abstract}



\section{Introduction and notations}\label{intr}

\subsection{Introduction}
The existence of interface conducting states is one of the most significant properties of topological insulators \cite{Kane2010,Khanikave2013}.  It originates in certain energy gaps caused by symmetry breaking in the bulk.  In the past two decades, there have been many efforts made to reveal the underlying mechanism of interface conducting states, mainly on the edge states caused by $\mcp\mct$ symmetry breaking \cite{Fefferman2012,Fefferman2016,Drouot2020,Zhu2023}.  However, physicists also find edge states in two-dimensional $\mcp\mct$ symmetric materials \cite{Hu2015}.  In this paper, we model such phenomena by analyzing the spectral properties of a domain wall modulated Schr\"odinger operator. 

One of the typical lattices whose deformed bulk has interface conducting states is the honeycomb lattice. The honeycomb lattice structure obsesses single Dirac cones at $K$ and $K'$ points and chiral edge states after time-reversal symmetry breaking \cite{Ablowitz2012,Fefferman2012,Fefferman2016,Drouot2019,Drouot2020}.  The most famous paradigm is graphene, a two-dimensional topological material \cite{Kane2010}. Around 2015, a new way to deform the honeycomb lattice has been reported \cite{Hu2015,Hu2018,Yves2017}. They considered the lattice in a supercell so that it has folding symmetry - caused by smaller periods than the supercell lattice. We call it a super honeycomb lattice in our previous work \cite{Cao2022}. This new highly symmetric structure has fourfold degeneracy and a double Dirac cone at the $\Gamma$ point. Deforming the super honeycomb lattice in two different directions makes the energy bands open a gap with different topologies near the $\Gamma$ point \cite{Cao2022,Miao2023}. Physicists found a pair of gapped pseudospin edge states when connecting such two types of deformed materials, which are analogs of the helical edge states \cite{Palmer2013}. The propagation of electromagnetic waves with frequencies between these two edge states is well confined near the interface  \cite{Barik2016,Smironova2019}. Despite the broad applications of this phenomenon, it is rarely studied rigorously. Motivated by this, we consider edge operators interpolating between two kinds of perturbed operators across a rational edge and try to establish the existence of two gapped edge states by rigorous analysis.  

Many models are used to develop analysis on edge states. Ammari and his collaborators studied the related mechanism on the subwavelength scale by considering Helmholtz problems \cite{Ammari2022A,Ammari2022B}. Bal and his collaborators have analyzed properties and topological descriptions of edge states in Dirac operators \cite{Bal2019,Bal2023}. Also, some numerical methods are introduced to associated problems \cite{Zhu2021,Guo2022}.

Among all the models, the one particle non-relativistic Schr\"odinger equation is one of the most effective models in illustrating edge conducting states \cite{Fefferman2016,Lee2019,Weinstein2020A,Drouot2020,Guo2022}. Fefferman and Weinstein have laid solid foundations of rigorous analysis on such equations in both $\mcp$ and $\mct$ breaking case \cite{Fefferman2012,Fefferman2014,Fefferman2016}. In this paper, we consider the following two-dimensional Schr\"odinger operator: $$\mch^{\delta}_{edge} = - \Delta + V(\bx) + \delta \eta(\delta \bm{l}_2 \cdot \bx) W(\bx),$$ with $V(\bx)$ a super honeycomb lattice potential, $\eta(\zeta)$ a domain function, and $W(\bx)$ a $\mcp\mct$ symmetry preseving perturbation; see section \ref{edge-operator}. The limiting perturbed bulk operators on two sides: $$\mch^{\delta}_{\pm} = - \Delta + V(\bx) \pm \delta  W(\bx)$$ are $\mcp\mct$ symmetric, but the folding symmetry is missing. Different from the analysis of $\mcp$ or $\mct$ breaking bulk operators \cite{Fefferman2016,Lee2019,Drouot2020}, we have to deal with the double Dirac cone \cite{Cao2022}, or two tangent single Dirac cones on the bands of the unperturbed operator $\mch_V = -\Delta + V(\bx)$, where a nontrivial second-order degeneracy is hidden; see section \ref{near-energy-appro} . This means that only in higher-order terms can we clearly see the interaction between the four branches when discussing edge states. Such behavior is deeply rooted in the $\mcp\mct$ symmetry preserving property.  In the time-reversal symmetry breaking case, a one-dimensional Dirac operator can be obtained as the effective model by asymptotic expansions. An edge state with energy near the Dirac point can be derived from its exponentially decaying zero-energy states \cite{Lee2019,Weinstein2020A,Drouot2020}. However, when it comes to the time reversal symmetric case, there are a pair of paralleling Dirac operators and therefore, two indistinguishable zero-energy states. Thus, the energy gap between two edge states is of higher order and deserves precise description; see section \ref{edge-multiscale}.

Another exciting perspective of understanding the edge states is the interplay between the symmetries and topology. Similar to how the Chern numbers characterize the topology of quantum materials, some topological indices are introduced to characterize the topology of bulk and edge Hamiltonians and the bulk-edge correspondence \cite{Ogata2020,Runder2013}. There exist results on some one-dimensional models \cite{Thiang2023,Thiang2022} and higher-dimensional Dirac operators \cite{Bal2019,Bal2023,Bal2023B}. {Different from a quantity of the whole band, in this paper, we provide topological intepretations concentrated on the $\Gamma$ point by calculating the local topological charge and analyzing the parities of the eigenstates of limiting bulk operators $\mch^{\delta}_{\pm}$ on two sides of the edge, which connects the symmetries and the topology more explicitly; see Section \ref{sec-topo}. }

This paper is organized as follows. Section \ref{rational-edge-operator} introduces the problem and our model -- the domain wall modulated edge operator. Results on super honeycomb lattice potentials and double Dirac cones are briefly reviewed in this section. Section \ref{near-energy-appro} establishes the approximations of the eigenstates near the double Dirac cone. Such near-energy approximations play essential roles in calculating the two gapped edge states asymptotically and rigorously proving their existence. The following three sections focus on the main conclusions obtained from our model. Section \ref{edge-multiscale} calculates two edge states explicitly by multiscale expansions. Section \ref{rigorous} proves the existence of two gapped edge states rigorously. {Section \ref{physical-intepretation} provides some physical interpretations. Subsection \ref{sec-index} is about the effective local topological charge. Subsection \ref{sec-topo} is about parties of eigenmodes. Section \ref{sec-num} gives numerical simulation of a typical example. }

\subsection{Notations}
\begin{itemize}
	\item { The lattice and dual lattice:} $\bf{U} = \Z\buu_1 \oplus \Z\buu_2$ denotes the parallelogram lattice in $\R^2$ expanding by $\buu_1$ and $\buu_2$. $\Omega = \big\{ \buu = c_1 \buu_1 + c_2\buu_2: c_1,c_2\in(-1/2,1/2) \big\}$ denotes its fundamental cell. ${\bf U}^* = \Z\bk_1 \oplus \Z\bk_2 $ denotes its dual lattice with $\bk_l\cdot\buu_j = 2\pi \delta_{l,j}$. $\Omega^* = \R^2/{\bf U}^*$ is the fundamental cell of its dual lattice and is called the Brillouin Zone.
     {
    \item Symmetry operators:
        \begin{enumerate}    
	\item $\frac{2}{3}\pi$ rotation operator $\mcr$:  $\mcr[f](\bx) = f(R^*\bx)$ with $R^*= \begin{pmatrix}  -\frac{1}{2} & -\frac{\sqrt{3}}{2}  \\  \frac{\sqrt{3}}{2} & -\frac{1}{2}   \end{pmatrix}$;
	\item reflection operator $\mcp$: $\mcp[f](\bx) = f(-\bx)$;
	\item time reversal operator $\mct$: $\mct[f](\bx) = \overline{f(\bx)}$.
    \end{enumerate} 

    \item The folding lattice and operator: when $\buu_2 = -R^*\buu_1$, $\bv_1$ and $\bv_2$ as following are the periods of the folding lattice.
   $$
     \bv_1 = \frac{1}{3}(2\buu_1 - \buu_2), \qq \bv_2 = \frac{1}{3}(\buu_1 + \buu_2).
   $$
   The associated folding operators are: $\mcv_l[f](\bx) = f(x+\bv_l)$.
    }
    
    \item {The rational edge: }$\bm{w}_1 = a_1\buu_1 + b_1 \buu_2$; $\bm{w}_2 = a_2\buu_1 + b_2 \buu_2$. $a_1,b_1,a_2,b_2$ are integers and $a_1b_2-a_2b_1=1$. $\bm{l}_1 = b_2\bk_1 - a_2 \bk_2$ and $\bm{l}_2 = -b_1\bk_1 + a_1 \bk_2$ satisfy {$\bm{w}_m\cdot\bm{l}_j =2\pi\delta_{m,j}$}. {$\bm{w}_1$ is the direction of the rational edge. $\bm{l}_2$ is its dual direction. } $\Tilde{\bm{l}}_1 = \bm{l}_1 - \frac{\bm{l}_1\cdot\bm{l_2}}{\Vert\bm{l}_2\Vert^2}{\bm{l}_2}$ is orthogonal to $\bm{l}_2$ and satisfies $\Tilde{\bm{l}}_1 \cdot \bm{w}_1 = 2\pi$.
      
	\item $\Omega_e = \big\{ \buu = c_1 \bm{w}_1 + c_2\bm{w}_2: c_1\in(-1/2,1/2), c_2\in\R  \big\}$. 
	\item There are three different inner products:
		\begin{enumerate}
			\item[1.] $\langle f(\bx),g(\bx)\rangle_{L^2(\Omega)} = \int_{\Omega} \overline{f(\bx)} g(\bx) d\bx $;
			\item[2.] $\langle f(\bx),g(\bx)\rangle_{L^2(\Omega_e)} = \int_{\Omega_e} \overline{f(\bx)} g(\bx) d\bx $;
			\item[3.] $\langle f(\zeta),g(\zeta)\rangle_{L^2(\R)} = \int_{\R} \overline{f(\zeta)} g(\zeta) d\zeta $.
		\end{enumerate}
    {The first one is mainly used in the bulk problem, and the last two are used in the edge problem.}
	\item There are two particular function spaces often used {corresponding to bulk and edge problem respectively}:
	$$\chi = \big\{ f \in L^2_{loc}(\R^2): f(\bx+\buu_l) = f(\bx) , l=1,2\big\};$$ $$\chi_e = \big\{ f \in L^2_{loc}(\R^2): f(\bx+\bm{w}_1) = f(\bx), \int_{\Omega_e} |f(\bx)|^2 d\bx < \infty \big\}.$$
	\item  Pauli matrices:
		\begin{equation*}
			\sigma_1 =\begin{pmatrix}
				0 &\hspace{0.2cm}  1\\
				1 &\hspace{0.2cm} 0
			\end{pmatrix};
			\quad \sigma_2 = \begin{pmatrix}
				0 &\hspace{0.1cm}  -i \\
				i &\hspace{0.1cm} 0
			\end{pmatrix}; \q 
			\sigma_3 = \begin{pmatrix}
				1 &\hspace{0.1cm}   0 \\
				0 &\hspace{0.1cm} -1
			\end{pmatrix}.
		\end{equation*}	
\end{itemize}

\section{Bulk and edge operators}\label{rational-edge-operator}

\begin{figure}[htbp]
    \centering
    \subfigure[]{\includegraphics[width=7.5cm]{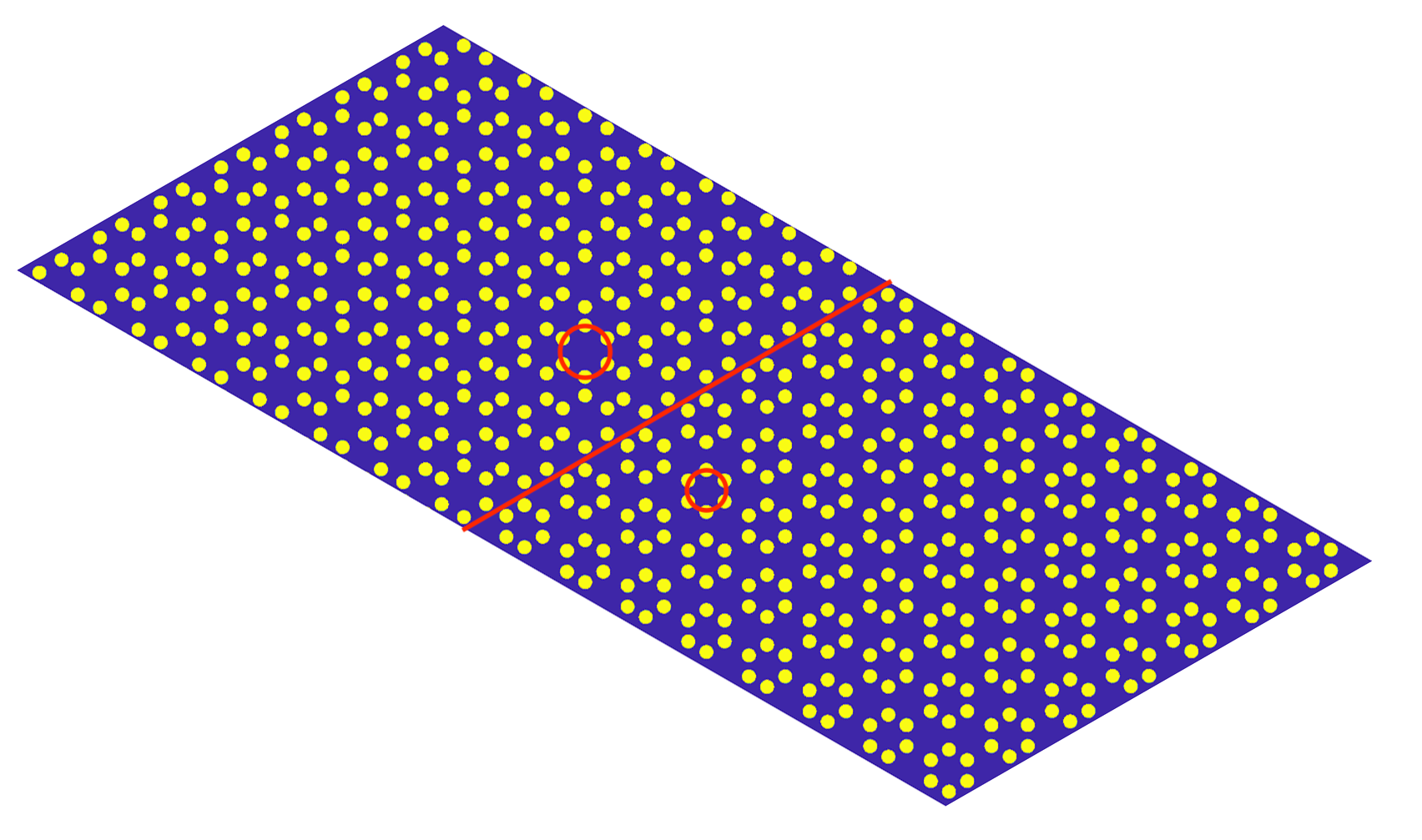}}
    \subfigure[]{\includegraphics[width =7cm]{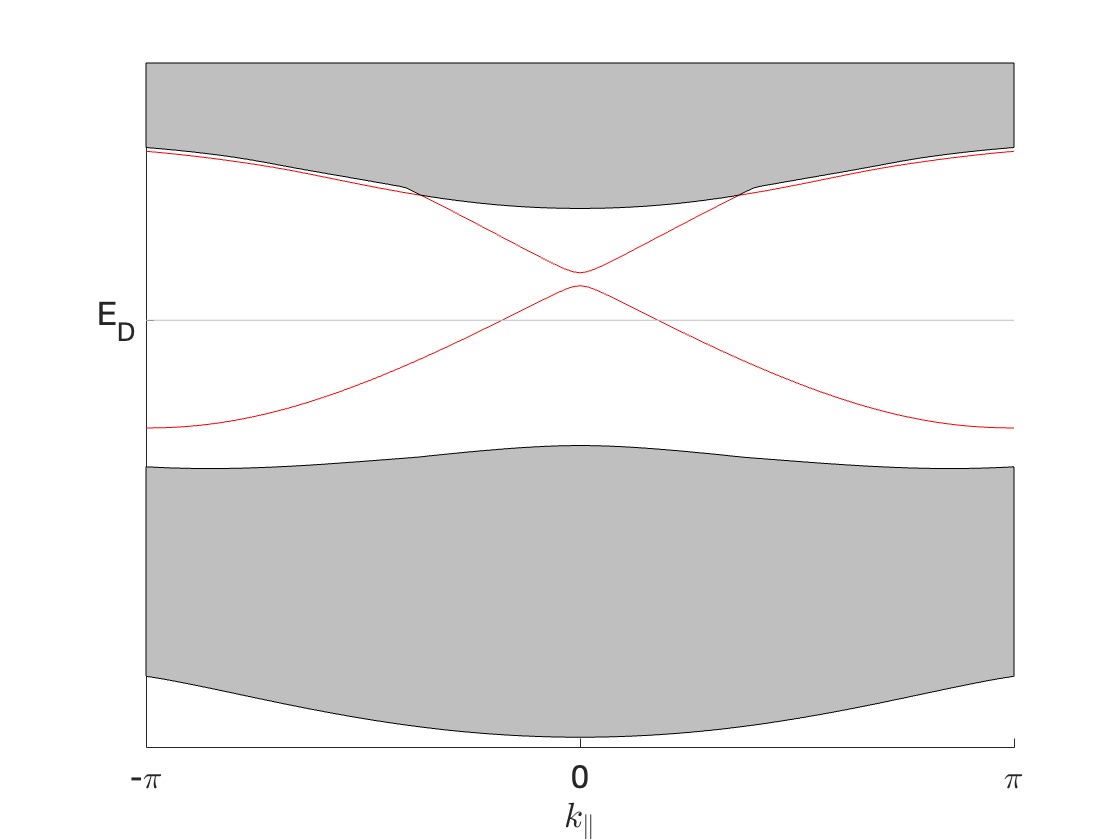}}
    \caption{Numerical simulations of edge states curves of a limiting domain wall model. (a) the figure of the piecewise constant domain wall potential. On one side of the edge, the hexagons of the super honeycomb lattice are shrunk, and on the other side, they are expanded. (b) the figure of the edge states energy curves along $k_{\parallel}\Tilde{\bm{l}}_1$. $E_{_D}$ is the Dirac point's energy. The parts of two red curves near the $\Gamma$ point correspond to two edge states.}
    \label{figure-energy-curve}
\end{figure}

The original bulk Hamiltonians should have $C_6$, $\mcp\mct$, and folding symmetries. We have studied such Schr\"odinger operators in the form of $\mch_V = -\Delta + V(\bx)$, where V(\bx) is a super honeycomb lattice potential as in Definition \ref{def-superHoneycomb}. The perturbed bulk operator is $\mch^{\delta} = - \Delta + V(\bx) + \delta W(\bx)$, where W(\bx) preserves $C_6$ and $\mcp\mct$ symmetries but destroys the folding symmetry \cite{Cao2022}. There should be two gapped edge states when gluing two kinds of perturbed bulk operators along certain edges, as shown in Figure \ref{figure-energy-curve}. In this paper, we talk about an asymptotic model - the domain wall modulated Schr\"odinger operator: $$\mch^{\delta}_{edge} = - \Delta + V(\bx) + \delta \eta(\delta \bm{l}_2 \cdot \bx)W(\bx),$$ where $\eta(\zeta)$ is a domain wall function as in Definition \ref{def-domainWall}. $\mch^{\delta}_{edge}$ is a slow interpolation between $\mch^{\delta}_{\pm} =  - \Delta + V(\bx) \pm \delta W(\bx)$ across a rational edge $\R \bm{w}_1$. This section briefly reviews bulk operators, including their symmetries and the degeneracy on their bands at the $\Gamma$ point, and introduces the edge operator.

\subsection{Super honeycomb lattice potentials and the folding symmetry}

In this subsection, we review the symmetries of super honeycomb lattice potentials. It originates in viewing the honeycomb lattice in a supercell. Thus, the super honeycomb lattice possesses a folding symmetry, which results in fourfold degeneracy at the $\Gamma$ point on energy bands.

Fefferman and Weinstein have summarized the structures of the honeycomb lattice in their paper \cite{Fefferman2012}. It is such a periodic structure in $\R^2$:
\begin{itemize}
	\item $\frac{2}{3}\pi$-rotation symmetric;
	\item $\mcp\mct$ (parity and time reversal) symmetric.	
\end{itemize}

The corresponding transformations in $L^2_{loc}(\R^2)$ are:
\begin{itemize}
	\item $\frac{2}{3}\pi$ rotation operator $\mcr$:  $\mcr[f](\bx) = f(R^*\bx)$ with $R^*= \begin{pmatrix}  -\frac{1}{2} & -\frac{\sqrt{3}}{2}  \\  \frac{\sqrt{3}}{2} & -\frac{1}{2}   \end{pmatrix}$;
	\item reflection operator $\mcp$: $\mcp[f](\bx) = f(-\bx)$;
	\item time reversal operator $\mct$: $\mct[f](\bx) = \overline{f(\bx)}$.
\end{itemize} 

\begin{definition}\label{def-honeycomb}{\bf (Honeycomb lattice potentials)}
    $V(\bx)\in C^{\infty}(\R^2)$ is called a honeycomb lattice potential, if
\begin{enumerate} 
    \item[1.] $V(\bx)$ is doubly periodic with periods $\buu_1$ and $\buu_2$, where $\buu_2 = -R^*\buu_1 $;
    \item[2.] $\mcr[V](\bx) = V(\bx)$;
    \item[3.] $\mcp[V](\bx) = V(\bx)$ and $\mct[V](\bx) = V(\bx)$.
\end{enumerate}
\end{definition}

Consider the parallelogram lattice $\bf{U}=\Z \buu_1 \oplus \Z\buu_2$. The corresponding unit cell is
\begin{equation}\label{eqn-unit-cell}
    \Omega = \big{\{} \buu = c_1 \buu_1 + c_2 \buu_2, \q c_1,c_2\in(-1/2,1/2) \big{\}}.
\end{equation}
Denote its dual lattice and dual unit cell $\bf{U}^* = \Z \bk_1 \oplus \Z \bk_2$ and $\Omega^* = \R^2/\bf{U}^*$, where $\bk_j$ satisfies $\buu_l\cdot \bk_j = 2\pi \delta_{l,j}$. 

\begin{figure}[htbp]
    \centering
    \subfigure[]{\includegraphics[width=5.0cm]{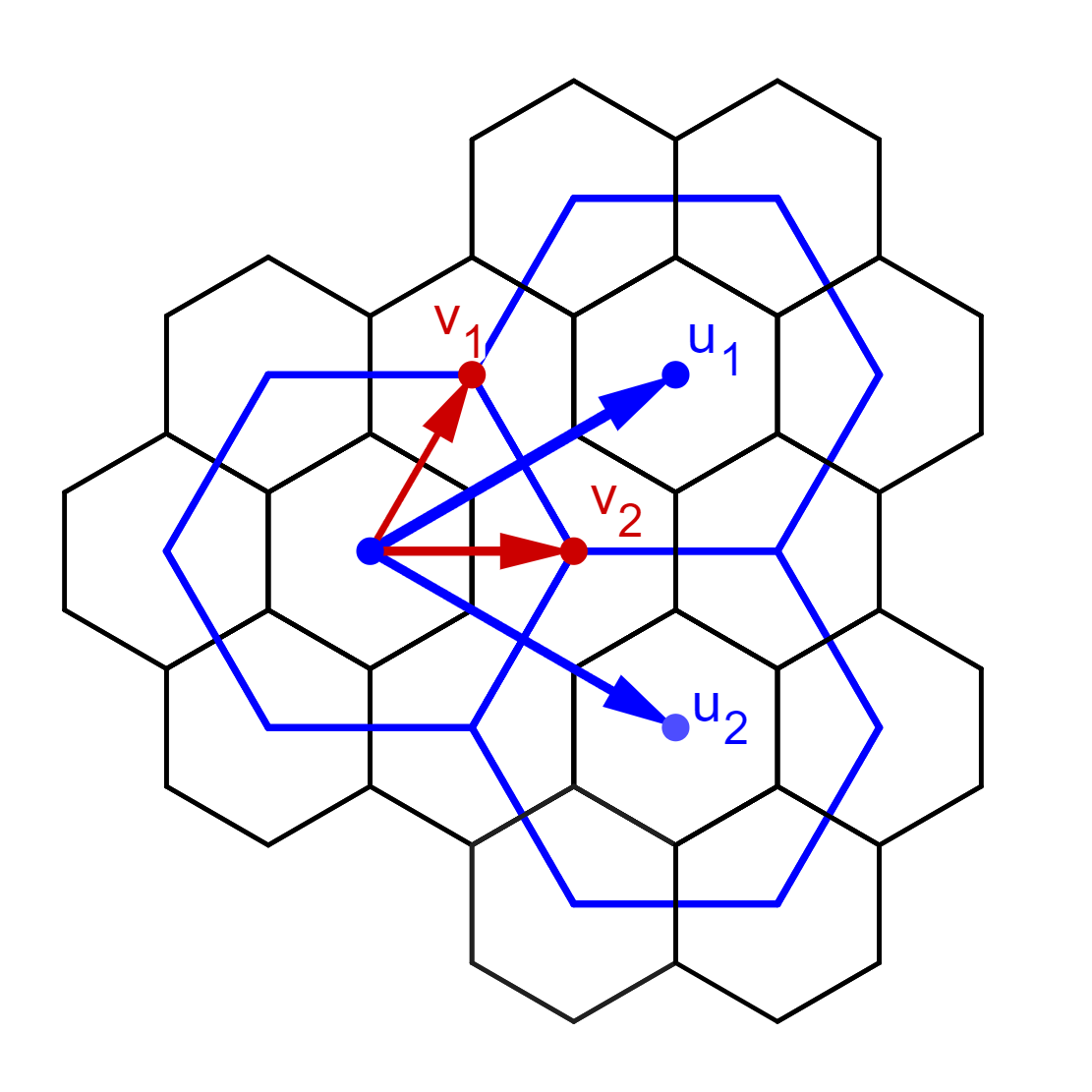}}\qq
    \subfigure[]{\includegraphics[width = 5.0cm]{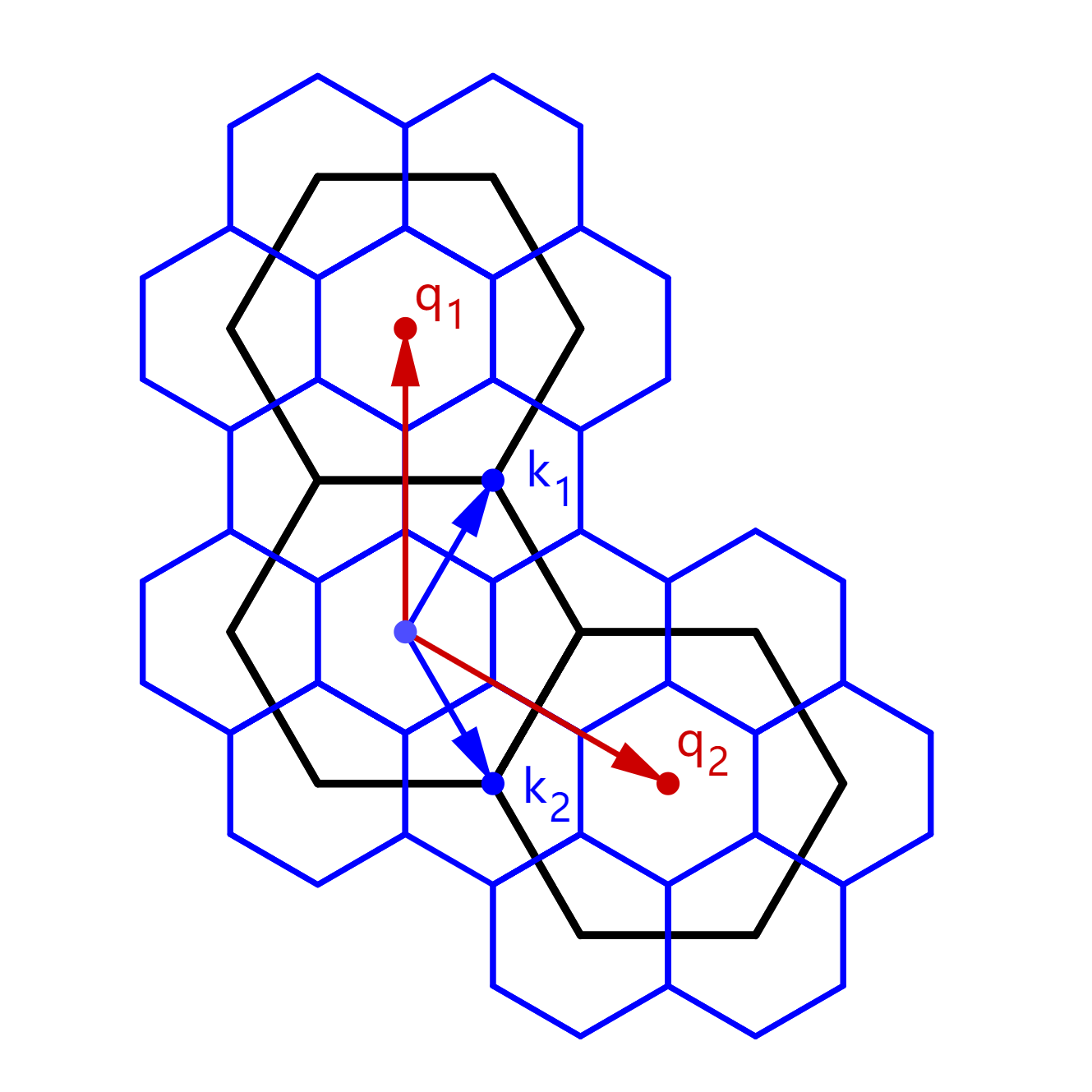}}

    \caption{(a) The lattice, and (b) the dual lattice. The blue lattice corresponds to a honeycomb lattice, with periods $\buu_1$ and $\buu_2$ and dual periods $\bk_1$ and $\bk_2$. The black lattice corresponds to the associated super honeycomb lattice, with periods $\bv_1$ and $\bv_2$ and dual periods $\bq_1$ and $\bq_2$.}\label{fig-lattice}
\end{figure}

The relation and differences between honeycomb lattice and super honeycomb lattice are shown explicitly in Figure \ref{fig-lattice}. Note that the parallelogram lattice discussed above and the hexagonal lattice in the pictures are equivalent. Compared with the honeycomb lattice potentials, the super honeycomb lattice potential obsesses smaller periods and, therefore, a folding symmetry. Namely, the smaller periods are
\begin{equation}\label{eqn-bv}
    \bv_1 = \frac{1}{3}(2\buu_1 - \buu_2), \qq \bv_2 = \frac{1}{3}(\buu_1 + \buu_2).
\end{equation}
For simiplicity, denote translation operators: $\mcv_l[f](\bx) = f(x+\bv_l)$. Let $\bq_j$ be the dual vectors of $\bv_l$, i.e., $\bq_j\cdot\bv_l = 2\pi \delta_{j,l}$. 

The super honeycomb lattice potential is defined below.

\begin{definition}\label{def-superHoneycomb}{\bf (Super honeycomb lattice potentials)}
    A honeycomb lattice potential $V(\bx)\in C^{\infty}(\R^2)$ is called a super honeycomb lattice potential if 
    \begin{enumerate}
        \item[4.] V(\bx) is $\bv_1$ and $\bv_2$ periodic, where $\bv_1$ and $\bv_2$ are as in (\ref{eqn-bv}) and the following condition holds:
        \begin{equation}\label{eqn-nondegeneracy}
           \frac{1}{|\Omega|}\int_{\Omega} e^{-i \bq_1 \cdot \by}V(\by) d \by \neq 0 .
        \end{equation}
    \end{enumerate}
\end{definition}

\begin{remark}
	The condition (\ref{eqn-nondegeneracy}) guarantees that the lowest Fourier element of $V(\bx)$ does not vanish, which prevents $V(\bx)$ from being a constant or possessing smaller periods. 
\end{remark}

\subsection{Bulk operators}

The bulk operators $\mch_{\pm}^{\delta} = -\Delta + V(\bx)\pm\delta W(\bx)$ on two sides of the edge are deformed from the highly symmetric operator $\mch_V = -\Delta + V(\bx)$, where $V(\bx)$ is a super honeycomb lattice potential. In this subsection, we give two conclusions about $\mch_V$ and $\mch^{\delta} = -\Delta + V(\bx)+\delta W(\bx)$ \cite{Cao2022}.

Generally, for a $\buu_1$ and $\buu_2$ doubly periodic elliptic operator $\mch$, by Floquet-Bloch theorem, its spectrum can be decomposed into momentum space $\Omega^*$: 
\begin{equation*}    
	\sigma_{L^2(\R^2)}(\mch) = \bigcup_{\bk \in \Omega^*} \sigma_{L^2_{\bk}(\R^2/ {\bf U})}(\mch|_{L^2_{\bk}(\R^2/ {\bf U})}).
\end{equation*}
Here the $\bk$-momentum function space is the Hilbert space $$L^2_{\bk}(\R^2/ {\bf U}) = \big{\{} f \in L^2_{loc}(\R^2): f(\bx+\buu) = e^{i\bk\cdot\buu}f(\bx), \bx\in \R^2, \buu \in{\bf U};  f(\bx)\in L^2(\Omega) \big{\}}$$
equipped with the inner product:
\begin{equation}
    \langle f(\bx),g(\bx) \rangle_{L^2(\Omega)} = \int_{\Omega} \overline{f(\bx)}g(\bx) d\bx.
\end{equation}
For our bulk operators, $\sigma_{L^2_{\bk}(\R^2/ {\bf U})}(\mch|_{L^2_{\bk}(\R^2/ {\bf U})})$ is a lower bounded discrete set:
\begin{equation}
    E_1(\bk) \leq E_2(\bk) \leq E_3(\bk) \leq ...
\end{equation}
$\mcs _n = \big\{\bigl(\bk,E_n(\bk)\bigr): \bk\in \Omega^* \big \}$ is the $n^{th}$ energy surface of $\mch$. Dirac cone describes the phenomenon of conical touch between energy surfaces. It is a kind of degeneracy and singularity on energy surfaces.

The first conclusion is that $\mch_V$ has fourfold degeneracy at the $\Gamma$ point and a double Dirac cone for general super honeycomb lattice potential $V(\bx)$ due to the symmetry. The key is that the space of $\buu_1$ and $\buu_2$ periodic functions $\chi = L^2_{\mathbf{0}}(\R^2/{\bf U})$ has the decomposition:
\begin{equation}
    \chi = \bigoplus_{\xi_1 ,\xi_2 =1 ,\tau,\overline{\tau}}\chi_{\xi_1,\xi_2} ,\q \tau = e^{\frac{2}{3}\pi i}.
\end{equation}
$\chi_{\xi_1,\xi_2}$ is the intersection of the characteristic subspace of $\mcv_1$ with eigenvalue $\xi_1$ and the characteristic subspace of $\mcr$ with eigenvalue $\xi_2$. Because any super honeycomb lattice potential $V(\bx)$ is in $\chi_{1,1}$, $\mch_V$ has the rotation and folding symmetries. Thus, each $\chi_{\xi_1,\xi_2}$ is an invariant subspace of $\mch_V$. The following theorem shows the existence of fourfold degeneracy and its relationship with the above function space decomposition.
 
\begin{theorem}\label{thm-doublecone}
    {\bf (Fourfold degeneracy at the $\Gamma$ point)} The following is true for energy surfaces of $\mch^{(\epsilon)}=-\Delta + \epsilon V(\bx)$ with $V(\bx)$ a super honeycomb lattice potential for all $\epsilon \in \R \setminus A$, where $A$ is a discrete subset of $\R$ :
    \begin{enumerate}
        \item there exists $n_*\in \N$ and $E_{_D} \in \R$ such that $\big\{ \mcs_{n_*+j}\big\}_{j=1,2,3,4}$ intersect at the $\Gamma$ point: 
        \begin{equation}\label{eqn-fourFold}
            E_{n_*}(\mathbf{0}) < E_{_D} = E_{n_*+1}(\mathbf{0})= E_{n_*+2}(\mathbf{0})  = E_{n_*+3}(\mathbf{0}) = E_{n_*+4}(\mathbf{0}) <  E_{n_*+5}(\mathbf{0});
        \end{equation}
        \item there exists $\phi_1(\bx)\in \chi_{\tau,\tau}$ normalized, such that
        \begin{equation*}
            \Ker(\mch^{(\epsilon)}-E_{_D}) = \Span\big{\{}\phi_1(\bx), \phi_2(\bx), \phi_3(\bx), \phi_4(\bx)\big{\}};
        \end{equation*}
        and the four eigenstates satisfy: 
        \begin{equation}\label{eqn-eigenFunc}
        	\begin{aligned}
        		\phi_1(\bx) \in \chi_{\tau,\tau}, \qq& \q\phi_2(\bx) = \overline{\phi_1(-\bx)} \in \chi_{\tau, \overline{\tau}},\\
        		\phi_3(\bx) = \phi_1(-\bx) \in \chi_{\overline{\tau}, \tau},\q&\q\phi_4(\bx) = \overline{\phi_1(\bx)} \in \chi_{\overline{\tau}, \overline{\tau}}.
        	\end{aligned}
        \end{equation}   \end{enumerate}  
\end{theorem}

Equation (\ref{eqn-fourFold}) indicates the existence of a fourfold degeneracy at the $\Gamma$ point. (\ref{eqn-eigenFunc}) provides insight into the four eigenfunctions corresponding to this degeneracy, revealing that they can be related through symmetries. We will give the result of the double Dirac cone later after we calculate the second-order bifurcation terms; see section \ref{sec-2nd-birfurcation}. 

The second conclusion is that the energy surfaces of $\mch^{\delta}$ open a gap of order $O(\delta)$ at the $\Gamma$ point when W(\bx) is a folding symmetry breaking potential as in the following definition.
\begin{definition}\label{def-perturbation}{\bf (Folding symmetry breaking potentials)}
A folding symmetry breaking potential $W(\bx)$ is a function in $C^{\infty}(\R^2)$ satisfying:
	\begin{itemize}
	\item $\mcp\mct$ symmetry preserving : $W(\bx)$ is a honeycomb lattice potential;
	\item folding symmetry breaking: $W(\bx) \in {\chi_{\tau,1}} \oplus {\chi_{\bar{\tau},1}}$, which means $W(\bx)$ is orthogonal to the space of super honeycomb lattice potentials so that it is ``purely" folding symmetry breaking. 
\end{itemize}
\end{definition}

The second conclusion is stated precisely as follows.

\begin{theorem}\label{thm-perturbation} {\bf (Local gap under folding symmetry breaking perturbations)}
    Let $\mch_V = -\Delta + V(\bx)$ be an operator possessing a fourfold degeneracy at the $\Gamma$ point as in Theorem \ref{thm-doublecone}. Assume that $W(\bx)$ is a folding symmetry breaking potential as above in Definition \ref{def-perturbation} and satisfies the non-degeneracy condition:
    \begin{equation}\label{eqn-cSharp}
     c_{\sharp} = \langle \phi_1(\bx) , W(\bx)\phi_1(-\bx) \rangle_{L^2(\Omega)} \neq 0.
     \end{equation}
     Then the energy surfaces $\big\{\mcs_{n_*+j} \big\}_{j=1,2,3,4}$ of perturbed operator $\mch^{\delta} = -\Delta + V(\bx) + \delta W(\bx)$ will open a gap of $O(|\delta|)$ near the $\Gamma$ point for $\delta$ sufficiently small.
\end{theorem}

\begin{remark}\label{remark-c}	
The $c_{\sharp}$ is real and stays invariant under phase transformation \cite{Cao2022}. Nonzero $c_{\sharp}$ guarantees that the gap is of order $O(\delta)$ exactly. 
\end{remark}

\subsection{Domain wall modulated edge operator}\label{edge-operator}
In this paper, we consider two kinds of bulks $\mch^{\delta}_{\pm} = -\Delta + V(\bx) \pm \delta W(\bx)$ connected along a rational edge $\R \bm{w}_1$. Here $\bm{w}_1 = a_1\buu_1 +b_1\buu_2$, where $a_1$ and $b_1$ are relatively prime integers. Then there exist relatively prime integers $a_2,b_2$ such that $a_1b_2 - a_2b_1 = 1$. Let  $\bm{w}_2 = a_2\buu_1 + b_2\buu_2$. Take $\bm{l}_1 = b_2\bk_1 - a_2\bk_2$ and $\bm{l}_2 = -b_1\bk_1 + a_1\bk_2$. Then $\bm{w}_j\cdot\bm{l}_l = 2\pi\delta_{j,l}$. Note that $\bm{w}_1$ and $\bm{w}_2$ expands a cell equivalent to $\Omega$ and $\bm{l}_1$ and $\bm{l}_2$ expands a cell equivalent to $\Omega^*$. Let $\Omega_e$ denote    such an area around the edge:
$$\Omega_e = \{\bx \in \R^2 \mid \bx = p\bm{w}_1 +q \bm{w}_2,  p\in(-1/2,1/2), q\in\R \} .$$

For the rational edge $\R\bm{w}_1$, we use the following asymptotic domain wall modulated operator:
\begin{equation}
    \mch_{edge}^{\delta} = - \Delta + V(\bx) + \delta \eta(\delta \bm{l}_2\cdot \bx) W(\bx).
\end{equation}  
The limiting bulk operators $\mch^{\delta}_{\pm}$ have been introduced in the last subsection. The function $\eta(\delta \bm{l}_2\cdot \bx)$ is the domain wall modulation function. It is flat along the edge direction $\bm{w}_1$ and varies slowly in the direction $\bm{w}_2$.  $\zeta = \delta \bm{l}_2\cdot \bx$ is called the slow variable. The rigorous definition of a domain wall function $\eta(\zeta)$ is below.

\begin{definition}\label{def-domainWall}{\bf (Domain wall functions)}
    A smooth function $\eta(\zeta)\in C^{\infty}(\R)$ is called a domain wall function if $\eta(\zeta)$ tends to $\pm 1$ when $\zeta \to \pm\infty$ and $\eta(0)=0$. 
\end{definition}

Similar to that in bulk case, we have:
\begin{equation*}    
	\sigma_{L^2(\R^2)}(\mch_{edge}^{\delta}) = \bigcup_{k_\parallel \in [-\pi,\pi)} \sigma_{L^2_{k_\parallel}(\R^2/ \Z \bm{w}_1)}(\mch_{edge}^{\delta}|_{L^2_{k_\parallel}(\R^2/ \Z \bm{w}_1)}).
\end{equation*}
Here the function space is one-dimensional quasi-periodic: $$L^2_{k_\parallel}(\R^2/ \Z \bm{w}_1) = \big{\{} f \in L^2_{loc}(\R^2) :f(\bx + \bm{w}_1) = e^{ik_\parallel}f(\bx) ,\bx\in \R^2; f(\bx)\in L^2(\Omega_e)  \big{\}}.$$
The inner product on $L^2_{k_\parallel}(\R^2/ \Z \bm{w}_1)$ is :
\begin{equation}
	\langle f , g\rangle_{L^2(\Omega_e)} = \int_{\Omega_e} \overline{f(\bx)}g(\bx) d\bx.
\end{equation}

 We summarize our asymptotic model in the following definition.
 
\begin{definition}\label{def-edge-operator}{\bf (Folding symmetry breaking domain wall modulated edge operators)}
 	$\mch_{edge}^{\delta} = -\Delta + V(\bx)+\delta\eta(\delta\bm{l}_2 \cdot \bx) W(\bx)$ is called a folding symmetry breaking domain wall modulated edge operator if it is constructed as follows.
 	\begin{itemize}
 		\item The unperturbed bulk operator has a fourfold degeneracy: $\mch_V = -\Delta +V(\bx)$ with $V(\bx)$ a super honeycomb lattice potential is an operator possessing fourfold degeneracy $({\bf 0}, E_{_D})$ at the $\Gamma$ point on $(n_*+1)^{th}$ to $(n_*+4)^{th}$ bands as (\ref{eqn-fourFold}) and (\ref{eqn-eigenFunc}) in Theorem \ref{thm-doublecone}.  		
 		\item For $j=1,2,3,4$, the energy band $E_{n_*+j}(\bk)$ of $\mch_V$ satisfies the non-fold condition:\begin{equation}\label{eqn-non-fold}
          		E_{n_*+j}(\lambda \bm{l}_2) = E_{_D} \Leftrightarrow \text{there exists } m,n \in \Z, \text{such that } \lambda \bm{l}_2 = m\bk_1 + n\bk_2.
    	 	 \end{equation} 	
    	\item The folding symmetry is broken but the $\mcp\mct$ symmetry is preserved: $W(\bx)$ is a folding symmetry breaking potential as in Definition \ref{def-perturbation}.
 		\item The domain wall is rational: $\eta(\zeta)$ is a domain wall function as in Definition \ref{def-domainWall} and $\bm{l}_2 = -b_1\bk_1 + a_1\bk_2$ for some co-prime $b_1$ and $a_1$.
    	\end{itemize} 
 \end{definition}

\begin{remark}
	 By the continuity of the spectral band, the condition (\ref{eqn-non-fold}) means that $E_{_D}$ are the maximum or minimum of the lower or upper bands of the double Dirac cone along the $\bm{l}_2$ direction. This is a typical case when the four intersecting bands can open an energy gap along the $\bm{l}_2$ direction across the $\Gamma$ point under small folding symmetry breaking perturbations.
 \end{remark}
 
 Our aim is to solve the following eigenvalue problem near $k_{\parallel}=0$:
\begin{equation}\label{eqn-edge-eigen-problem}
    \mch_{edge}^{\delta} \psi(\bx, k_{\parallel}) = \mce(k_{\parallel})\psi(\bx, k_{\parallel}) ,\q  \psi(\bx, k_{\parallel})  \in L^2_{k_\parallel}(\R^2/ \Z \bm{w}_1).
\end{equation}
The following sections contribute to proving the existence of two gapped edge states near $k_{\parallel}=0$ and characterizing these two edge states precisely.

\section{Near-energy approximation}\label{near-energy-appro}

This section focuses on detailed information on the near-energy approximations -- the approximations of energies and eigenstates of $\mch_V$ with energy near $E_{_D}$, which helps to approximate the near-energy components when proving the existence of two gapped edges states in the following sections. More accurate results than the previous ones are needed because the first-order bifurcation matrix is still degenerate \cite{Cao2022}. For the four branches of energy surfaces intersecting at the $\Gamma$ point, the upper two do not separate at order $O(\Vert\bk\Vert)$, nor do the lower two. The second-order bifurcation matrix is calculated in this section so that the near-energy approximation along the direction $\bm{l}_2$ can be given precisely.

\subsection{Order $O(\Vert\bk\Vert^2)$ bifurcation matrix}\label{sec-2nd-birfurcation}

This subsection discusses the bifurcation matrices for any $\bk$ sufficiently small to estimate the near-energy eigenstates.

The eigenvalue problem of $\mch_V$ on $L^2_{\bk}(\R^2/ {\bf U})$:
\begin{equation*}
	\begin{aligned}
		\mch_V \phi(\bx;\bk) &= E(\bk)\phi(\bx;\bk),\\
		\phi(\bx;\bk) = e&^{i\bk\cdot\bx} p(\bx),\quad p(\bx)  \in \chi;
	\end{aligned}
\end{equation*}
is equivalent to 
\begin{equation*}
		\mch_V(\bk) p(\bx;\bk) = E(\bk) p(\bx;\bk),\quad p(\bx) \in \chi,
\end{equation*}
where $\mch_V(\bk) = -(\nabla + i\bk)^2 + V(\bx)$.

We only consider the near-energy solution:
\begin{equation*}
	E(\bk) = E_{_D} + \mu, \quad p(\bx;\bk) =   {\Phi}(\bx) ^{\mathrm{T}}\bf{P}(\bk) + \psi(\bx;\bk).
\end{equation*}
where $\bf{P}(\bk)$ $ = (p_1(\bk),p_2(\bk), p_3(\bk), p_4(\bk))^{\mathrm{T}}$, $\psi(\bx;\bk)\in \Ker(\mch_V-E_{_D})^{\perp}$, the corrector $\mu$ and $\psi(\bx;\bk)$ are small, and
\begin{equation}\label{eqn-PHI}
    {\Phi}(\bx) = (\phi_1(\bx),\phi_2(\bx), \phi_3(\bx), \phi_4(\bx))^{\mathrm{T}}.
\end{equation}Here $\{\phi_j(\bx)\}_{j}$ span $\Ker(\mch_V-E_{_D})$ as in (\ref{eqn-eigenFunc}). Substitute these into the eigenvalue problem and rearrange it into:
\begin{equation}\label{eqn-LSR}
	(\mch_V-E_{_D})\psi(\bx;\bk) = (\mu + 2i \bk \cdot \nabla - \Vert \bk\Vert ^2) \bigl(  {\Phi}(\bx) ^{\mathrm{T}}{\bf{P}}(\bk) + \psi(\bx;\bk)\bigr).
\end{equation}

Thus, the corrector $\psi(\bx;\bk)$ is the solution to the problem:
\begin{equation*}
	\begin{aligned}
		\bigl(I-(\mch_V-E_{_D})^{-1} \mcq_{\perp}(2i\bk \cdot \nabla & + \Vert \bk \Vert^2 + \mu)\bigr) \psi(\bx;\bk)
		 \\
   & = (\mch_V-E_{_D})^{-1} \mcq_{\perp} 2i\bk \cdot \nabla {\Phi}(\bx) ^{\mathrm{T}}\bf{P}(\bk),
	\end{aligned}
\end{equation*} 
where $\mcq_{\perp}$ is the projection map to the orthogonal complement space of $\Ker(\mch_V-E_{_D})$ and
\begin{equation}\label{eqn-inverse}
	(\mch_V -E{_D})^{-1}: \mcq_{\perp}\chi \to \mcq_{\perp}H^1(\R^2/ {\bf U}).
\end{equation}
Here $H^1(\R^2/ {\bf U})$ is the limitation of $\chi$ in $H^1(\R^2)$. Therefore, 
\begin{equation*}
	\psi(\bx;\bk) = \bigl((\mch_V-E_{_D})^{-1} \mcq_{\perp} 2i\bk \cdot \nabla {\Phi}(\bx) ^{\mathrm{T}} + O(\Vert\bk\Vert^2)\bigr) \bf{P}(\bk).
\end{equation*}

Besides, from (\ref{eqn-LSR}), we know that:
\begin{equation*}
	\langle  \phi_l(\bx),  (\mu + 2i \bk \cdot \nabla - \Vert \bk\Vert ^2) \bigl(  {\Phi}(\bx) ^{\mathrm{T}}{\bf{P}}(\bk) + \psi(\bx;\bk)\bigr)\rangle = 0.
\end{equation*}

Based on all these, solving the original eigenvalue problem can be reduced to solving possible $\bf{P}(\bk)$ in $(\mu I+B(\bk)) {\bf{P}}(\bk) =\bf{0}$, with\begin{equation*}
	B(\bk) = B_1(\bk)+ B_2(\bk) + O(\Vert \bk\Vert)^3.
\end{equation*}
Here 
\begin{equation*}
	\begin{aligned}
		B_1(\bk) &= \biggl( \big{\langle} \phi_l(\bx), 2i\bk\cdot\nabla \phi_j(\bx)\big{\rangle}_{L^2(\Omega)} \biggr) _{l,j} ;\\
		B_2(\bk) &= \biggl( \big{\langle} \phi_l(\bx),  \bigl(2i\bk \cdot \nabla (\mch_V-E_D)^{-1} \mcq_{\perp}  2i\bk \cdot \nabla  -\Vert\bk\Vert^2 \bigr)\phi_j(\bx)\big{\rangle}_{L^2(\Omega)} \biggr) _{l,j}.
	\end{aligned}
\end{equation*}
$B(\bk)$ represents the bifurcation matrix. \begin{enumerate}
	\item First-order bifurcation matrix $B_1(\bk)$.

	Taking advatange of the symmetry relations between $\phi_j(\bx)$ in (\ref{eqn-eigenFunc}), we can obtain \cite{Cao2022}
	\begin{equation*}
		B_1(\bk) = \begin{pmatrix}
			0 & 2i\bk \cdot \bm{v}_{\sharp} & 0 &0\\
			  \overline{2i\bk \cdot\bm{v}_{\sharp}} & 0 & 0& 0\\
			  0 & 0 & 0 & -2i\bk \cdot \bm{v}_{\sharp}\\
			  0 & 0 &  -\overline{2i\bk \cdot\bm{v}_{\sharp}} & 0
		\end{pmatrix}.
	\end{equation*}
	Here $\bm{v}_{\sharp}$ is such a quantity:
	\begin{equation}\label{eqn-v-sharp}
		 \bm{v}_{\sharp} = \langle \phi_1(\bx) ,\nabla \overline{\phi_1(-\bx)} \rangle_{L^2(\Omega)} = \frac{v_{_F}}{2}\begin{pmatrix}
        1 \\ i
    \end{pmatrix}e^{i\theta_{\sharp}}.
	\end{equation}		
	\begin{remark}
	The eigenvalues of $b(\bk)$ are $\lambda_{\pm} = \pm |2i\bk\cdot \bm{v}_{\sharp}| = \pm v_{_F}\Vert \bk\Vert $. Both $\lambda_{\pm}$ are of multiplicity two, which means that the upper two bands near the fourfold degenerate Dirac point are tangent to each other and so do the lower two bands. To distinguish the tangent bands, we have to calculate higher-order approximations. The Fermi velocity $v_{_F}$ is the slope of the double Dirac cone. It is invariant under phase transformation of $\phi_1(\bx)$. The corresponding eigenvectors are determined by $\theta_{\sharp}$, which naturally changes with the phase transformation of $\phi_1(\bx)$. 
		
	\end{remark}	
		
  \item Second-order bifurcation matrix $B_2(\bk)$. 
	
	The second-order bifurcation matrix is Hermitian. Thus, it is enough to discuss the upper part of
	$$\biggl( \big{\langle} \phi_l(\bx),  \bigl(2i\bk \cdot \nabla (\mch_V-E_D)^{-1} \mcq_{\perp}  2i\bk \cdot \nabla \bigr)\phi_j(\bx)\big{\rangle}_{L^2(\Omega)} \biggr) _{l,j}.$$
	 First, due to
	\begin{equation*}
		\begin{aligned}
			& \langle \phi_l(\bx), 2i\bk \cdot \nabla (\mch_V-E_D)^{-1} \mcq_{\perp}  2i\bk \cdot \nabla  \phi_j(\bx) \rangle_{L^2(\Omega)}
			\\ = & \langle \mcq_{\perp} 2i\bk \cdot 
			\nabla \phi_l(\bx), (\mch_V-E_D)^{-1}\mcq_{\perp}2i\bk \cdot 
			\nabla \phi_j(\bx) \rangle_{L^2(\Omega)},
		\end{aligned}
	\end{equation*}
	and the fact that $(\mch_V-E_D)^{-1}$ is an elliptic and Hermitian operator on the periodic function space $\mcq_{\perp}L^2(\Omega)$, $\langle \phi_l(\bx), ...  \phi_l(\bx) \rangle_{L^2(\Omega)}$ are real for all $l$. Besides, using the symmetries between the eigenfunctions in Theorem \ref{thm-doublecone}, $\langle \phi_l(\bx), ...  \phi_l(\bx) \rangle_{L^2(\Omega)}$ are the same for all $l$. This means the diagonal elements of the two-order bifurcation matrix are just real multiples of the identity. For other elements, because the translation operator $\mcv_1$ is a unitary operator and commutative with both $\nabla$ and $(\mch_V-E_D)^{-1}$ on $L^2(\Omega)$, we have 
	\begin{equation*}
	\begin{aligned}
		 \langle \phi_l(\bx), ...  \phi_j(\bx) \rangle_{L^2(\Omega)}
		=  \langle \mcv_1\phi_l(\bx), ...\mcv_1\phi_j(\bx) \rangle_{L^2(\Omega)}.
	\end{aligned}		
	\end{equation*}
	Because all $\phi_l(\bx)$ are eigenfunctions of $\mcv_1$, it turns out that the two-order bifurcation matrix is quasi-diagonal:
	\begin{equation*}
	\begin{aligned}
		 0 =& \langle \phi_1(\bx), ...  \phi_3(\bx) \rangle_{L^2(\Omega)} = \langle \phi_1(\bx), ... \phi_4(\bx) \rangle_{L^2(\Omega)}
		\\ = &\langle \phi_2(\bx), ... \phi_3(\bx) \rangle_{L^2(\Omega)} = \langle \phi_2(\bx), ... \phi_4(\bx) \rangle_{L^2(\Omega)}.
	\end{aligned}
	\end{equation*}

	Therefore, the second order bifurcation matrix $B_2(\bk)$ is equal to:
	\begin{equation*}
		\begin{pmatrix}
			m(\bk) -\Vert \bk \Vert^2 & b(\bk) & 0 & 0\\
			\overline{b(\bk)} & m(\bk) -\Vert \bk \Vert^2 & 0 & 0 \\
			0 & 0 & m(\bk) -\Vert \bk \Vert^2 & b(\bk) \\
			 0 &  0 &  \overline{b(\bk)} & m(\bk) -\Vert \bk \Vert^2
		\end{pmatrix}.
	\end{equation*}
	 where
	 \begin{equation}\label{eqn-m}
	 	 		m(\bk) = \langle \phi_1(\bx), 2i\bk \cdot \nabla (\mch_V-E_D)^{-1} \mcq_{\perp}  2i\bk \cdot \nabla   \phi_1(\bx) \rangle _{L^2(\Omega)};
	 \end{equation}
	 \begin{equation}\label{eqn-b1}
	 		b(\bk)  = \langle \phi_1(\bx), 2i\bk \cdot \nabla (\mch_V-E_D)^{-1} \mcq_{\perp}  2i\bk \cdot \nabla   \overline{\phi_1(-\bx)} \rangle _{L^2(\Omega)}.
	 \end{equation} 
	 \begin{remark}
	 	$m(\bk)$ is real. The diagonal part $\bigl(m(\bk) -\Vert \bk \Vert^2\bigr)I$ of this matrix does no contribution to bifurcation. The term resulting in second-order bifurcation of the upper or lower two bands is $b(\bk)$. The form of $b(\bk)$ is calculated explicitly in Appendix \ref{appen}.
	 \end{remark}
\end{enumerate}

\begin{remark}\label{remark-v}
	Both the first-order and second-order bifurcation matrices are quasi-diagonal. This is because the eigenvalue problems of super honeycomb lattice potential on $\bigcup\limits_{*\in\{1,\tau,\bar{\tau}\}} \chi_{\tau,*}$ and $\bigcup\limits_{*\in\{1,\tau,\bar{\tau}\}} \chi_{\bar{\tau},*}$ are decoupled. 
\end{remark}

Taking use of these discussions, we finally obtain second-order accurate near-energy approximations of the four branches intersecting at the $\Gamma$ point as the following theorem:

\begin{theorem}\label{thm-second-order-cone}{\bf(Double Dirac cone at the $\Gamma$ point)}
	Let $\mch_V = -\Delta + V(\bx)$ be a Schr\"odinger operator with a super honeycomb lattice potential $V(\bx)$. Assume that it has a fourfold degeneracy at the $\Gamma$ point as in (\ref{eqn-fourFold}) and the corresponding four eigenstates $\big\{\phi_l(\bx)\big\}$ satisfy the symmetry condition (\ref{eqn-eigenFunc}). Let $\bm{v}_{\sharp}$, $m(\bk)$, and $b(\bk)$ denote the same terms as in (\ref{eqn-v-sharp}) , (\ref{eqn-m}) and (\ref{eqn-b1}). Suppose that $\bm{v}_{\sharp}$ satisfies the non-degeneracy condition:
	\begin{equation}\label{eqn-v-nondegeneracy}
		v_{_F} = \Vert \bm{v}_{\sharp}\Vert  = \Vert\langle \phi_1(\bx), \nabla \overline{\phi_1(-\bx)}\rangle_{L^2(\Omega)} \Vert \neq 0.
	\end{equation}
	Then, the intersecting four branches behave conically near the $\Gamma$ point:
	\begin{equation}
      \begin{aligned}\label{eqn-E-222}
            E_{n_*+1}(\bk) = E_{_D}  - \check{\mu}_1(\bk)+ \Vert\bk\Vert^2 -m(\bk) + O(\Vert \bk\Vert^3 )),\\
            E_{n_*+2}(\bk) = E_{_D}  - \check{\mu}_2(\bk)+ \Vert\bk\Vert^2 -m(\bk)+ O(\Vert \bk\Vert^3 )),\\
            E_{n_*+3}(\bk) = E_{_D}  + \check{\mu}_2(\bk)+ \Vert\bk\Vert^2 -m(\bk)+ O(\Vert \bk\Vert^3 )),\\
            E_{n_*+4}(\bk) = E_{_D}  + \check{\mu}_1(\bk) +\Vert\bk\Vert^2 -m(\bk)+ O(\Vert \bk\Vert^3 )).
     \end{aligned}
\end{equation}
The term corresponding to the first two order bifurcation is:
\begin{equation*}
	\begin{aligned}
		\check \mu_1(\bk) = \max \bigl(| 2i\bk\cdot \bm{v}_{\sharp}+    b(\bk)|, | 2i\bk\cdot \bm{v}_{\sharp}-    b(\bk)|\bigr),\\
		\check \mu_2(\bk) = \min \bigl(| 2i\bk\cdot \bm{v}_{\sharp}+    b(\bk)|, | 2i\bk\cdot \bm{v}_{\sharp}-    b(\bk)|\bigr).
	\end{aligned}
\end{equation*} 
\end{theorem}

\begin{remark}
	The condition (\ref{eqn-v-nondegeneracy}) guarantees that  the intersecting four branches $\big\{ \mcs_{n_*+j}(\bk)\big\}_{j=1,2,3,4}$ do not behave too flat near the $\Gamma$ point. 
\end{remark}

This theorem characterizes the double Dirac cone up to second-order accuracy near the fourfold degenerate point. 

Suppose that $| 2i\bk\cdot \bm{v}_{\sharp}+ b(\bk)|>| 2i\bk\cdot \bm{v}_{\sharp}- b(\bk)|$. The related approximations of four branches of ${\bf{P}}(\bk)$ are:
\begin{equation*}
	{\bf{P}}^{n_*+1}(\bk) = \begin{pmatrix}
		|2i\bk\cdot \bm{v}_{\sharp}+ b(\bk)| \vspace{0.2cm} \\
		\overline{2i\bk\cdot \bm{v}_{\sharp}+ b(\bk)} \\ 0\\0
	\end{pmatrix}+ O(\Vert \bk \Vert ^3);
\end{equation*}
\begin{equation*}
     {\bf{P}}^{n_*+2}(\bk) = \begin{pmatrix}
		 0\\0 \\- |2i\bk\cdot \bm{v}_{\sharp}- b(\bk)| \vspace{0.2cm} \\
		\overline{2i\bk\cdot \bm{v}_{\sharp}- b(\bk)} 
	\end{pmatrix}+ O(\Vert \bk \Vert ^3);
\end{equation*}
\begin{equation*}
	{\bf{P}}^{n_*+3}(\bk) = \begin{pmatrix}
		0 \\ 0 \\|2i\bk\cdot \bm{v}_{\sharp}- b(\bk)| \vspace{0.2cm} \\
		\overline{2i\bk\cdot \bm{v}_{\sharp}- b(\bk)} 
	\end{pmatrix}+ O(\Vert \bk \Vert ^3); 
\end{equation*}
\begin{equation*}
	{\bf{P}}^{n_*+4}(\bk) = \begin{pmatrix}
		 -|2i\bk\cdot \bm{v}_{\sharp}+ b(\bk)| \vspace{0.2cm} \\
		\overline{2i\bk\cdot \bm{v}_{\sharp}+ b(\bk)} \\ 0\\0
	\end{pmatrix}+ O(\Vert \bk \Vert ^3).
\end{equation*}
Similar results can be obtained for $| 2i\bk\cdot \bm{v}_{\sharp}+ b(\bk)|<| 2i\bk\cdot \bm{v}_{\sharp}- b(\bk)|$ with these $P^{j}(\bk)$ reordered.

\subsection{Near-energy approximation along certain direction}\label{near-energy}

This subsection gives the near-energy approximation along the $\bm{l}_2$ direction, which can be written in an analytic form.

For the given direction $\bm{l}_2$, we can choose a typical $e^{i\theta^*}\phi_1(\bx)$, which is also in $\chi_{\tau,\tau}$, to replace the original $\phi_1(\bx)$, such that the new $\bm{v}_{\sharp}$ has the following property \cite{Drouot2020}:
\begin{equation}\label{eqn-vSharp-l2}
     - 2i\br \cdot \bm{v}_{\sharp} = \frac{v_{_F}}{\Vert \bm{l}_2\Vert}\big{(}\br\cdot \bm{l}_2 - \Det[\br, \bm{l}_2] i\big{)},\q \br \in \R^2.
\end{equation}
In the following content throughout this paper, we fix the $\phi_1(\bx)$ such that $\bm{v}_{\sharp}$ always take such a simple form (\ref{eqn-vSharp-l2}).
 
\begin{proposition}\label{prop-near-appro}{\bf (Near-energy approximation along $\bm{l}_2$ direction)}
    Let $V(\bx)$ be a super honeycomb lattice potential and $\mch_V = -\Delta +V(\bx)$ be a Schr\"odinger operator as in Theorem \ref{thm-second-order-cone}. Assume  (\ref{eqn-vSharp-l2}) is true for $\bm{v}_{\sharp}$ and $b(\bm{l}_2)\neq0$ where $b(\bk)$ is as in (\ref{eqn-b1}). Then there exists a $\lambda_0>0$ such that for all $\vert \lambda \vert < \lambda_0$ the following is true:
    \begin{enumerate}
        \item $\{E_{n_*+j}(\lambda \bm{l}_2)\}_{j=1,2,3,4}$ are equivalent to such four real analytic functions intersecting at $\lambda =0 $:
        \begin{equation}
            \begin{aligned}
                \theta_1(\lambda) = E_{_D} - v_{_F}\Vert \bm{l}_2 \Vert \lambda + r_1(\lambda)\lambda^2,\\
                \theta_2(\lambda) = E_{_D} - v_{_F}\Vert \bm{l}_2 \Vert \lambda + r_2(\lambda)\lambda^2,\\
                \theta_3(\lambda) = E_{_D} + v_{_F}\Vert \bm{l}_2 \Vert \lambda + r_3(\lambda)\lambda^2,\\
                \theta_4(\lambda) = E_{_D} + v_{_F}\Vert \bm{l}_2 \Vert \lambda + r_4(\lambda)\lambda^2,
            \end{aligned}
        \end{equation}
        with $|r_j(\lambda)| < C$, where $C$ is a positive constant independent of $\lambda$.
        \item Corresponding orthonormal eigen modes in $\Ker\bigl(\mch_V-\theta_j(\lambda)\bigr)$ in $L^2_{\lambda\bm{l}_2}(\R^2/{\bf U})$ can be chosen real analytically dependent on $\lambda$:
        \begin{equation}
            \begin{aligned}
                \Theta_1(\bx;\lambda) =\frac{1}{\sqrt{2}}e^{i\lambda \bm{l}_2 \cdot \bx}\bigl(\phi_1(\bx)-\phi_2(\bx)\bigr)+ R_1(\bx;\lambda),\\
                \Theta_2(\bx;\lambda) =\frac{1}{\sqrt{2}} e^{i\lambda \bm{l}_2 \cdot \bx}\bigl(\phi_3(\bx)+\phi_4(\bx)\bigr)+ R_2(\bx;\lambda),\\
                \Theta_3(\bx;\lambda) =\frac{1}{\sqrt{2}} e^{i\lambda \bm{l}_2 \cdot \bx}\bigl(\phi_3(\bx)-\phi_4(\bx)\bigr)+ R_4(\bx;\lambda),\\
                \Theta_4(\bx;\lambda) = \frac{1}{\sqrt{2}}e^{i\lambda \bm{l}_2 \cdot \bx}\bigl(\phi_1(\bx)+\phi_2(\bx)\bigr)+ R_3(\bx;\lambda).
            \end{aligned}
        \end{equation}
        Here $R_j(\bx;\lambda) = O_{H^2(\R^2/{\bf U})}(\lambda )$.
    \end{enumerate}
\end{proposition}

\begin{remark}\label{remark-effective-mass}
    {The quantity \( b(\bk) \) in (\ref{eqn-b1}) and \( m(\bk) \) in (\ref{eqn-m}) are closely related to the second-order derivative of the energy, which corresponds to the effective mass in physical literature. For example, one of the four analytic functions \( \theta_j \) can be expressed as:
   \[
   \theta_{j_0}(\lambda) = E_{_D} - \left| v_{_F} \Vert \bm{l}_2 \Vert \lambda - b(\bm{l}_2)\lambda^2 \right| + (\Vert \bm{l}_2 \Vert^2 - m(\bm{l}_2)) \lambda^2 + O(\lambda^3).
   \]
   Combined with (\ref{eqn-E-222}), this shows that \( b(\bk) \) governs the effective mass difference between the upper or lower pair bands. The condition \( b(\bm{l}_2) \neq 0 \) ensures that these pairs split at order \( O(\Vert \bk \Vert^2) \).}
\end{remark}

\Proof
	With (\ref{eqn-vSharp-l2}), the expressions of $\theta_j(\lambda)$ and $\Theta(\bx;\lambda)$ can be obtained directly from the discussion in the last subsection. We first solve out four $\theta_j(\lambda)$ real analytic dependent on $\lambda$ from $\det(\theta I + B(\lambda \bm{l}_2)) =0$ where $B(\bk)$ is the bifurcation matrix in the last subsection, and then find the four corresponding $\Theta_j(\bx; \lambda)$ which are also real analytically dependent on $\lambda$ \cite{Fefferman2016}.     
\qed
\section{Multiscale expansions of edge states}\label{edge-multiscale}

In this paper, the proof of the existence of two gapped edge states of the domain wall modulated operator $\mch_{edge}^{\delta}$ can be divided into two main parts: calculating the main terms of energies and eigenstates by multiscale expansions first and estimating the remaining terms following the classical method as in the paper of Fefferman and Weinstein \cite{Fefferman2016} then. This section focuses on calculating the main terms. The most important conclusion in this section is the existence of the gap of order $O(\delta^2)$ under natural assumptions on second-order bifurcation terms. Specifically, we can expand the energies of these two edge states at $k_{\parallel}=0$ as $$\mce_j(0) = E_{_D} + \delta \mce^{(1)}_j(0) + \delta^2 \mce^{(2)}_j(0) + O(\delta^3).$$ The two main conclusions in this part are:
\begin{itemize}
	\item the linear terms $\mce^{(1)}_j(0)$ are both zero for $j=1,2$, and they correspond to a two-dimensional zero energy eigenspace of an operator which can be diagonalized into two Dirac operators;
	\item the quadratic terms $\mce^{(2)}_j(0)$ bifurcate when second-order bifurcation term (\ref{eqn-main-1}) is nonzero, which coincides with the non-degeneracy condition of order $O(\Vert\bk\Vert^2)$ bifurcation matrix; see the last section.
\end{itemize}

Let us take a more convinient direction $\tilde{\bm{l}}_1 = \bm{l}_1 - \frac{ \bm{l}_1 \cdot \bm{l}_2 }{\Vert \bm{l}_2 \Vert ^2}\bm{l}_2$, which is orthogonal to $\bm{l}_2$ and satisfies $ \tilde{\bm{l}}_1 \cdot \bm{w}_1 = 2\pi $. Let $k_{\parallel} = \delta s \tilde{\bm{l}}_1\cdot \bm{w}_1$, and denote: 
\begin{equation*}
    \check{L}_s^2(\R^2 / \Z \bm{w}_1) = \big\{ f\in L^2(\Omega_e) \q  | \q f(\bx) = e^{i \delta s \tilde{\bm{l}}_1\cdot\bx} p(\bx),\q  p(\bx + \bm{w}_1) = p(\bx)  \big\}.
\end{equation*} In this section, we aim to solve the following eigenvalue problem by multiscale expansions:
\begin{equation}\label{eqn-s-problem}
    \mch_{edge}^{\delta} \psi(\bx; s) = \mce(s)\psi(\bx, s) ,\q  \psi(\bx; s)  \in  \check{L}_s^2(\R^2 / \Z \bm{w}_1).
\end{equation}

For simplicity, we denote the space of $\bm{w}_1$-periodic functions:
\begin{equation}\label{eqn-chi-e}
    \chi_e =  \check{L}_0^2(\R^2/\Z\bm{w}_1) = \{ f\in L^2(\Omega_e) \q  | \q f(\bx + \bm{w}_1) = f(\bx)  \}.
\end{equation}

For $\delta$ small, $s$ near zero, and $\mce(s)$ near $E_{_D}$, using the slow variable $\zeta = \delta \bm{l}_2 \cdot \bx $ and associated multiscale solution $\psi(\bx, \zeta; s)$. Expand the solution in powers of $\delta$:
\begin{equation}
    \begin{aligned}
        \psi(\bx, \zeta; s) & = \psi^{(0)}(\bx, \zeta;s) + \delta \psi^{(1)}(\bx, \zeta;s) + \delta ^2 \psi^{(2)}(\bx, \zeta; s) + ... , \\
        \mce(s) & = E_{_D} + \delta \mce^{(1)}(s) + \delta ^2 \mce^{(2)}(s) + ...  
    \end{aligned}
\end{equation}
Then, substituting the powers into the equation and grouping the terms by order in $\delta$ to obtain equations for these $\psi^{(j)}(\bx, \zeta,s)$ and $ \mce^{(j)}$.

At order $O(1)$, it is: 
\begin{equation}\label{eqn-order-delta-0}
    \bigl(\mch_V - E_{_D} \bigr) \psi^{(0)}(\bx, \zeta; s) =0,
\end{equation}
where $\mch_V = -\Delta_{\bx} + V(\bx) $.

At order $O(\delta)$, it is:
\begin{equation}\label{eqn-order-delta-1} 
         \bigl( \mch_V - E_{_D}\bigr) \psi^{(1)}(\bx,\zeta;s)  = \mcl_1(s) \psi^{(0)}(\bx, \zeta; s),
\end{equation}
where $\mcl_1(s) =  2 \bigl(is\tilde{\bm{l}}_1  + \partial_{\zeta} \bm{l}_2 \bigr)\cdot 
\nabla_{\bx}  - \eta(\zeta) W(\bx) + \mce^{(1)}(s) $.

At order $O(\delta^2)$, it is:
\begin{equation}\label{eqn-order-delta-2}
    \bigl( \mch_V - E_{_D}\bigr) \psi^{(2)}(\bx,\zeta;s) = \mcl_1(s) \psi^{(1)}(\bx, \zeta; s) + \mcl_2(s)\psi^{(0)}(\bx, \zeta; s),
\end{equation}
where $\mcl_2(s) = \bigl( is\tilde{\bm{l}}_1  + \partial_\zeta \bm{l}_2 \bigr)^2 + \mce^{(2)}(s)$.

At order $O(\delta^n)$, $n\geq 3$, it is:
\begin{equation}
    \begin{aligned}\label{eqn-order-delta-n}
        & \bigl( \mch_V - E_{_D}\bigr) \psi^{(n)}(\bx,\zeta;s) \\ = &\mcl_1(s) \psi^{(n-1)}(\bx, \zeta; s) + \mcl_2(s) \psi^{(n-2)}(\bx, \zeta; s) + \sum_{j=3}^n \mce^{(j)}(s)\psi^{(n-j)}(\bx, \zeta; s).
    \end{aligned}
\end{equation}

\subsection{Order $O(\delta)$ terms}\label{sec-order-delta}

For equation (\ref{eqn-order-delta-0}), we can use the ansatz 
\begin{equation}\label{eqn-ansatz-0}
    \psi^{(0)}(\bx, \zeta; s)  = {\Phi}(\bx) ^{\mathrm{T}} \bm{\alpha}(\zeta;s),
\end{equation}
 where $\bm{\alpha}(\zeta;s) = (\alpha_1(\zeta;s),\alpha_2(\zeta;s), \alpha_3(\zeta;s), \alpha_4(\zeta;s))^{\mathrm{T}}$, and
 $\Phi(\bx)$ as in (\ref{eqn-PHI}). Substituting the ansatz into (\ref{eqn-order-delta-1}), by the solvable condition for $(\mch_V - E_{_D})$, the right-hand side of (\ref{eqn-order-delta-1}) should be orthogonal to $\Ker(\mch_V- E_{_D})$:
\begin{equation}
    \langle \Phi(\bx), \mcl_1(s) \Phi(\bx)^{\mathrm{T}} \rangle _{L^2(\Omega)}\bm{\alpha}(\zeta,s) = 0;
\end{equation}
to be specific:
\begin{equation}\label{eqn-order-delta}
    \begin{aligned}
        0 = & \quad2  \langle \Phi(\bx),  \nabla_{\bx} {\Phi}(\bx)^{\mathrm{T}} \rangle  _{L^2(\Omega)}\cdot \biggl(is\tilde{\bm{l}}_1 \bm{\alpha}(\zeta;s)+ \partial_{\zeta}\bm{\alpha}(\zeta;s) \bm{l}_2 \biggr) \\+ & \mce^{(1)}\langle \Phi(\bx), {\Phi}(\bx)^{\mathrm{T}}   \rangle _{L^2(\Omega)}\bm{\alpha}(\zeta; s) 
          -  \langle \Phi(\bx), W(\bx){\Phi}(\bx)^{\mathrm{T}} \rangle  _{L^2(\Omega)} \eta(\zeta)\bm{\alpha}(\zeta; s).
    \end{aligned}
\end{equation}

With the help of (\ref{eqn-vSharp-l2}), (\ref{eqn-cSharp}), and (\ref{eqn-eigenFunc}) the symmetric relations between $\phi_j(\bx)$, (\ref{eqn-order-delta}) is such an equation:
\begin{equation}\label{eqn-effective-Dirac-1}
    (\mcd(s) - \mce^{(1)}(s) I ) \bm{\alpha}(\zeta;s) = 0, 
\end{equation}
where
\begin{equation}
    \mcd(s) = \Det [\Tilde{\bm{l}}_1, \bm{l}_2] \frac{v_{_F}}{\Vert \bm{l}_2 \Vert } s \sigma_3 \otimes \sigma _2 + \frac{1}{i} v_{_F}\Vert \bm{l}_2 \Vert \sigma_3 \otimes \sigma _1  \partial_{\zeta}   + c_{\sharp} \eta(\zeta) \sigma_1 \otimes I.
\end{equation}
$\sigma _j$ are pauli matrices.

Use the following orthogonal transformation $Q$ to decouple the original 4-dimensional problem (\ref{eqn-effective-Dirac-1}) into two 2-dimensional problems:
\begin{equation}\label{eqn-trans-Q}
    Q = \begin{pmatrix}
        \frac{\sqrt{2}}{2}  &   & \frac{\sqrt{2}}{2}  &  \\
        &  \frac{\sqrt{2}}{2}  &   &  \frac{\sqrt{2}}{2}\\
        \frac{\sqrt{2}}{2} &   & -\frac{\sqrt{2}}{2}  &  \\
        & -\frac{\sqrt{2}}{2} &   & \frac{\sqrt{2}}{2}\\
    \end{pmatrix}.
\end{equation}

Then the effective Dirac operator for order $O(\delta)$ is
\begin{equation}
    \Tilde{ \mcd}(s) = Q^T\mcd(s)Q = \diag(\mcd_1(s), \mcd_2(s)),
\end{equation}
where
\begin{equation}
    \begin{aligned}
        \mcd_1(s) = Det [\Tilde{\bm{l}}_1, \bm{l}_2] \frac{v_{_F}}{\Vert \bm{l}_2 \Vert } s \sigma _2 + \frac{1}{i} v_{_F}\Vert \bm{l}_2 \Vert \sigma _1  \partial_{\zeta}   + c_{\sharp} \eta(\zeta) \sigma_3,\\
        \mcd_2(s) = Det [\Tilde{\bm{l}}_1, \bm{l}_2] \frac{v_{_F}}{\Vert \bm{l}_2 \Vert } s \sigma _2 + \frac{1}{i} v_{_F}\Vert \bm{l}_2 \Vert \sigma _1  \partial_{\zeta}   - c_{\sharp} \eta(\zeta) \sigma_3.
    \end{aligned}
\end{equation}

Thus, solving eigenvalue problem (\ref{eqn-effective-Dirac-1}) is equivalent to finding eigenvalues of $\mcd_1(s)$ and $\mcd_2(s)$. Note that $\mcd_2(s) = -\overline{\mcd_1(s)}$. We have the following proposition for the eigenvalue and eigenfunctions for $\mcd_1(s)$ and $\mcd_2(s)$. Its proof is simple, and we omit it here.

\begin{proposition}\label{prop-order-delta}{\bf (First-order approximation of the edge states)} $\mcd_1(s)$ has a simple eigenvalue 
    \begin{equation*}
        \mu(s) = \sgn(c_{\sharp}) \Det [\Tilde{\bm{l}}_1, \bm{l}_2] \frac{v_{_F}}{\Vert \bm{l}_1 \Vert} s
    \end{equation*}
    with the normalized eigenstate:
    \begin{equation*}
        \bm{d}(\zeta) = \begin{pmatrix}
            \sgn(c_{\sharp}) \\  i 
        \end{pmatrix} \alpha_{\sharp}(\zeta) , 
    \end{equation*}
    where
    \begin{equation}\label{eqn-alpha-sharp}
    	\alpha_{\sharp}(\zeta)= c_{\alpha}\exp\biggl(-\frac{|c_{\sharp}|}{v_{_F}\Vert \bm{l}_2 \Vert} \int_0^{\zeta} \eta(t) dt\biggr),
    \end{equation}
    and $c_{\alpha}$ is a real normalization coefficient for $\bm{d}(\zeta)$. Similarly, $\mcd_2(s)$ has a simple eigenvalue $-\mu(s)$ with the eigenstate $\overline{\bm{d}(\zeta)}$.
\end{proposition}

We can deduce the following conclusions from this proposition.
\begin{enumerate}
    \item When $s\neq 0$, $\mcd(s)$ has two different eigenvalues $\mce^{(1)}_1(s)=\mu(s)$ and $\mce^{(1)}_2(s)=-\mu(s)$, and corresponding eigenstates:
    \begin{equation*}
        \bm{\alpha}^1(\zeta;s) = Q^T \begin{pmatrix}
            \bm{d}(\zeta) \\ 0 \\ 0
        \end{pmatrix} ; \qq \bm{\alpha}^2(\zeta;s) = Q^T \begin{pmatrix}
             0 \\ 0 \\ \overline{\bm{d}(\zeta)}
        \end{pmatrix} .
    \end{equation*}
    \item {\bf (The first-order degeneracy of the edge states' energies)} When $s=0$, $0$ is an eigenvalue of $\mcd(0)$ of multiplicity two with such a pair of eigenstates as a basis:
    \begin{equation}\label{eqn-alpha}
    \begin{aligned}
        \bm{\alpha}^1(\zeta;0) = \alpha_{\sharp}(\zeta) \begin{pmatrix}
            \sgn(c_{\sharp}) \\ 0 \\ 0 \\ -i
        \end{pmatrix} ,  \qquad \bm{\alpha}^2(\zeta;0) = \alpha_{\sharp}(\zeta) \begin{pmatrix}
            0 \\ \sgn(c_{\sharp}) \\-i \\ 0
        \end{pmatrix} ,
    \end{aligned}
    \end{equation}
     Note that we choose two special orthogonal $\bm{\alpha}^l(\zeta;0)$ in $\Ker (\mcd(0))$ here to simplify the calculation in the next subsection. 
\end{enumerate}

Thus, for $s\neq0$, we have two different solutions at order $O(\delta)$ as above and we can take 
$
    \psi^{(0)}_j(\bx,\zeta;s) = {\Phi}(\bx) ^{\mathrm{T}} \bm{\alpha}^j(\zeta;s)
$
for $j=1,2$ respectively.

But for $s=0$, $\mce^{(1)}(0) = 0 $ is of multiplicity two, and we can only distinguish the two eigenvalues and corresponding eigenstates at higher orders. Now we just set
\begin{equation}\label{eqn-delta-2-}
    \psi^{(0)}(\bx,\zeta;0) = c^1 {\Phi}(\bx) ^{\mathrm{T}} \bm{\alpha}^1(\zeta;0) + c^2 {\Phi}(\bx) ^{\mathrm{T}} \bm{\alpha}^2(\zeta;0)
\end{equation}
and try to solve out two groups of $(c^1,c^2)$ at the next order.  

\subsection{Order $O(\delta^2)$ terms}\label{multi-two}

If $\psi^{(0)}(\bx,\zeta;s)$ is solved out,  we can recursively use the ansatz 
 \begin{equation}\label{eqn-ansatz-1}
    \psi^{(1)}(\bx, \zeta; s)  = {\Phi}(\bx)  ^{\mathrm{T}} \bm{\beta}(\zeta,s)+ (\mch_V-E_{_D})^{-1}\mcl_1(s)\psi^{(0)}(\bx,\zeta;s).
 \end{equation}
 where $\bm{\beta}(\zeta;s) = (\beta_1(\zeta;s),\beta_2(\zeta;s), \beta_3(\zeta;s), \beta_4(\zeta;s))^{\mathrm{T}}$, and $(\mch_V-E_{_D})^{-1}$ is as in (\ref{eqn-inverse}).
 
For $s\neq 0$, the original multiscale eigenvalue problem is split into two different problems whose eigenvalue has the first-order term of $\mu(s)\delta$ and $-\mu(s)\delta$ respectively, and the rest work is just in a traditional style. At order $O(\delta^2)$, using the ansatz (\ref{eqn-ansatz-1}), we can obtain
\begin{equation*}
     \bigl( \mch_V - E_{_D}\bigr) \psi^{(2)}(\bx,\zeta;s) = \mcl_1(s) {\Phi}(\bx)  ^{\mathrm{T}} \bm{\beta}(\zeta;s) + \mcf(s)  \psi^{(0)}_j(\bx,\zeta;s),
\end{equation*}
where $\mcf(s) =  \mcl_1(s) (\mch_V - E_D)^{-1}\mcl_1(s) +  \mcl_2(s)$. Use the solvable condition of $(\mch_V-E_{_D})$ again to solve out $\bm{\beta}_j(\zeta; s)$ and construct ansatzes like (\ref{eqn-ansatz-1}) recursively for all the rest orders to get the multiscale expansions of two gapped eigenvalues $\mce_1(s)$ and $\mce_2(s)$ and associated eigenstates $\psi_1(\bx,\zeta; s)$ and $\psi_2(\bx,\zeta; s)$.

However, for $s=0$, at order $O(\delta^2)$, we can only obtain:
\begin{equation} \label{eqn-delta-2-1}
\begin{aligned}
    \bigl( \mch_V - E_{_D}\bigr) & \psi^{(2)}(\bx,\zeta;0) = \mcl_1(0)  {\Phi}(\bx)  ^{\mathrm{T}} \bm{\beta}_j(\zeta;0) \\ & + \mcf(0) \bigl( c^1 {\Phi}(\bx) ^{\mathrm{T}} \bm{\alpha}^1(\zeta;0) + c^2 {\Phi}(\bx) ^{\mathrm{T}} \bm{\alpha}^2(\zeta;0)\bigr).
\end{aligned}
\end{equation}
Similarly, (\ref{eqn-delta-2-1}) has to satisfy the solvable condition of $(\mch_V-E_{_D})$. Thus, the right-hand side of the equation is orthogonal to $\Ker(\mch_V-E_{_D})$, and we can obtain:
\begin{equation}\label{eqn-delta-2-2}
\begin{aligned}
     \mcd(0) \bm{\beta}(\zeta;0) =  - \big{\langle} \Phi(\bx),   \mcf(0)\bigl( c^1( {\Phi}(\bx) ^{\mathrm{T}} \bm{\alpha}^1(\zeta;0) + c^2 {\Phi}(\bx) ^{\mathrm{T}} \bm{\alpha}^2(\zeta;0)\bigr) \big{\rangle} _{L^2(\Omega)}.
\end{aligned}
\end{equation}
Because $\mcd(0)$ is self-adjoint, (\ref{eqn-delta-2-2}) has a solution if and only if the right-hand side is orthogonal to $\bm{\alpha}^l(\zeta;0)$ for $l=1,2$:
\begin{equation}\label{eqn-delta-2-3}
\begin{aligned}
      &\big{\langle} {\bm{\alpha}}^l(\zeta;0),  \big{\langle} \Phi(\bx),   \mcf(0)  {\Phi}(\bx) ^{\mathrm{T}} \bm{\alpha}^1(\zeta;0) \big{\rangle}_{L^2(\Omega)}\big{\rangle}_{L^2(\R)}c^1 \\
     +  & \big{\langle} {\bm{\alpha}}^l(\zeta;0),  \big{\langle} \Phi(\bx),   \mcf(0){\Phi}(\bx) ^{\mathrm{T}} \bm{\alpha}^2(\zeta;0) \big{\rangle}_{L^2(\Omega)}\big{\rangle}_{L^2(\R)} c^2 =0.
\end{aligned}
\end{equation}
The outer inner product is about $\zeta$ in $L^2(\R)$. This equation has nonzero solutions $(c^1, c^2)^{\mathrm{T}}$ if and only if $\Det B_e = 0 $, where 
\begin{equation}\label{eqn-delta-2-det}
    B_e = \big{\langle} \begin{pmatrix} {\bm{\alpha}}^1(\zeta;0)\\  {\bm{\alpha}}^2(\zeta;0) \end{pmatrix} , \langle \Phi(\bx),   \mcf(0){\Phi}(\bx) ^{\mathrm{T}} \begin{pmatrix} \bm{\alpha}^1(\zeta;0) & \bm{\alpha}^2(\zeta;0)  \end{pmatrix}  \rangle_{L^2(\Omega)}   \big{\rangle}_{L^2(\R)}.
\end{equation}

$B_e$ is the bifurcation matrix for the edge states' problem with the following property.

\begin{proposition}{\bf (Second-order approximation of the edge states)} $\Det B_e = 0$ has two real solutions $\mce^{(2)}_1(0)$ and $\mce^{(2)}_2(0)$.
\end{proposition}

\Proof
    From (\ref{eqn-delta-2-det}), we know that 
\begin{equation*}
    \begin{aligned}
        B_e = \mce^{(2)}(0)I + \big{\langle} \begin{pmatrix} {\bm{\alpha}}^1(\zeta;0)\\  {\bm{\alpha}}^2(\zeta;0) \end{pmatrix} , \langle \Phi(\bx),   \mcn {\Phi}(\bx) ^{\mathrm{T}} \begin{pmatrix} \bm{\alpha}^1(\zeta;0) & \bm{\alpha}^2(\zeta;0)  \end{pmatrix}  \rangle_{L^2(\Omega)}   \big{\rangle}_{L^2(\R)}.
    \end{aligned}
\end{equation*}

Here $\mcn = \big( 2\partial_{\zeta} \bm{l}_2 \cdot \nabla_{\bx} -\eta(\zeta) W(\bx) \big)(\mch_V-E_{_D})^{-1}\mcq_{\perp}\big( 2\partial_{\zeta} \bm{l}_2 \cdot \nabla_{\bx} -\eta(\zeta) W(\bx) \big) + (\partial_{\zeta} \bm{l}_2)^2$. Denote $\mcn_1 = 2\bm{l}_2\cdot \nabla (\mch_V-E_{_D})^{-1} \mcq_{\perp}  2\bm{l}_2 \cdot \nabla$, $\mcn_2  = 2\bm{l}_2 \cdot \nabla (\mch_V-E_{_D})^{-1}\mcq_{\perp}W(\bx)$, $\mcn_3  = W(\bx)(\mch_V-E_{_D})^{-1} \mcq_{\perp}2\bm{l}_2 \cdot \nabla$, and $\mcn_4 = W(\bx) (\mch_V-E_{_D})^{-1}\mcq_{\perp} W(\bx)$. Now, let us calculate the following terms first. 

\begin{enumerate}
	\item $\biggl(\langle \phi_l(\bx), \mcn_1 \phi_j(\bx) \rangle _{L^2(\Omega)}\biggr)_{l,j}:$ 
	 
	 	This can be calculated similarly with second-order bifurcation terms in section \ref{sec-2nd-birfurcation}. The final result is $$\biggl(\langle \phi_l(\bx), \mcn_1 \phi_j(\bx) \rangle _{L^2(\Omega)}\biggr)_{l,j} = m_1 I + I \otimes M_1.$$
	 	Here $M_1=\begin{pmatrix}
	 		0 & b_1 \\ \overline{b_1} & 0
	 	\end{pmatrix}$,  where $b_1 = \langle \phi_1(\bx), \mcn_1 \phi_2(\bx) \rangle _{L^2(\Omega)}$.	 	
	 		
	\item $\biggl( \langle 	\phi_l(\bx), \mcn_2\phi_j(\bx) \rangle_{L^2(\Omega)} \biggr)_{l,j}:$
 
		Because $\mcr$ is unitary, we have:
  \begin{equation*}
		\begin{aligned}
			& \langle \phi_l(\bx), \nabla (\mch_V-E_{_D})^{-1}\mcq_{\perp}W(\bx)\phi_j(\bx) \rangle_{L^2(\Omega)} 			\\
			= & \langle \mcr \bigl(\phi_l(\bx)\bigr), \mcr\bigl( \nabla (\mch_V-E_{_D})^{-1}\mcq_{\perp}W(\bx)\phi_j(\bx)\bigr) \rangle_{L^2(\Omega)} 			\\
			= & R^* \langle \mcr \bigl(\phi_l(\bx)\bigr),  \nabla (\mch_V-E_{_D})^{-1}\mcq_{\perp}W(\bx)\mcr\bigl(\phi_j(\bx)\bigr) \rangle_{L^2(\Omega)}.	
		\end{aligned}		
		\end{equation*}
		Since $\phi_j(\bx)$ and $\phi_l(\bx)$ are eigenfunctions of $\mcr$, the above quantity takes a nonzero vector only when it is an eigenvector of $R^*$ with eigenvalue $\tau$ or $\bar{\tau}$, which means $\phi_j(\bx)$ and $\phi_l(\bx)$ should correspond to different eigenvalues of $\mcr$. By the symmetries between the eigenfunctions in Theorem \ref{thm-doublecone}, we can get:
		$$\biggl( \langle 	\phi_l(\bx), \mcn_2\phi_j(\bx) \rangle_{L^2(\Omega)} \biggr)_{l,j} = \sigma_3 \otimes M_2 + \frac{1}{i}\sigma_2 \otimes M_3,$$	
	    where
	    \begin{equation*}
	    	M_2 = \begin{pmatrix}
	    		0 & b_2 \vspace{0.2cm} \\ 
	    		-\overline{b_2} & 0
	    	\end{pmatrix} \qq M_3 = \begin{pmatrix}
	    		0 &  -b_3 \vspace{0.2cm} \\ 
	    		\overline{b_3} & 0
	    	\end{pmatrix}.
	    \end{equation*}
	    $b_2= \langle \phi_1(\bx), \mcn_2\phi_2(\bx) \rangle_{L^2(\Omega)} $ and $b_3 = \langle \phi_1(\bx), \mcn_2\phi_4(\bx) \rangle_{L^2(\Omega)}$.

     \item $\biggl( \langle 	\phi_l(\bx), \mcn_3\phi_j(\bx) \rangle_{L^2(\Omega)} \biggr)_{l,j}:$
	
		Note that 
		\begin{equation*}
		\begin{aligned}
			& \langle 	\phi_l(\bx), \mcn_3 \phi_j(\bx) \rangle_{L^2(\Omega)} = \langle - \mcn_2	\phi_l(\bx), \phi_j(\bx) \rangle_{L^2(\Omega)} \\
			= &-\overline{\langle 	\phi_j(\bx), \mcn_2\phi_l(\bx) \rangle_{L^2(\Omega)}} .
		\end{aligned}
		\end{equation*}
	    Thus, $\biggl( \langle 	\phi_l(\bx), \mcn_3\phi_j(\bx) \rangle_{L^2(\Omega)} \biggr)_{l,j} = -\biggl( \langle 	\phi_l(\bx), \mcn_2\phi_j(\bx) \rangle_{L^2(\Omega)} \biggr)_{l,j}^{H}$.

		\item $\biggl( \langle 	\phi_l(\bx), \mcn_4\phi_j(\bx) \rangle_{L^2(\Omega)} \biggr)_{l,j}:$  
		 
		Again because $\mcr$ is unitary, $\langle 	\phi_l(\bx), \mcn_4\phi_j(\bx) \rangle_{L^2(\Omega)}$ takes a nonzero value only when $\phi_j(\bx)$ and $\phi_l(\bx)$ correspond to the same eigenvalue of $\mcr$. Denote $$m_4=\langle 	\phi_1(\bx), \mcn_4\phi_1(\bx) \rangle_{L^2(\Omega)};$$$$b_4   = \langle 	\phi_1(\bx), \mcn_4\phi_3(\bx) \rangle_{L^2(\Omega)}. $$ Then, due to the fact that $\mcn_4$ is Hermitian and the symmetries between $\phi_l(\bx)$, $m_4$ is real and $\langle 	\phi_l(\bx), \mcn_4\phi_l(\bx) \rangle_{L^2(\Omega)} = m_4$ for all $l$. $\mcp$ is unitary, too. Thus,
		\begin{equation*}
			\begin{aligned}
				b_4=& \langle \phi_1(\bx), \mcn_4\phi_3(\bx) \rangle_{L^2(\Omega)} = \langle \mcp \phi_1(\bx), \mcp \mcn_4\phi_3(\bx) \rangle_{L^2(\Omega)}\\
				= & \langle \phi_3(\bx), \mcn_4\phi_1(\bx) \rangle_{L^2(\Omega)} =  \langle \mcn_4\phi_3(\bx), \phi_1(\bx) \rangle_{L^2(\Omega)} 
			\end{aligned}
		\end{equation*}
		is real. The final result is:
		$$\biggl( \langle 	\phi_l(\bx), \mcn_4\phi_j(\bx) \rangle_{L^2(\Omega)} \biggr)_{l,j}= m_4 I + \sigma_1 \otimes M_4, $$
		where $M_4(\bk) = b_4I$.
\end{enumerate}

Based on these and the expressions of $\bm{\alpha}^l(\zeta)$, we can finally get that:
\begin{equation*}
	B_e = (\mce^{(2)}(0) + m_0) I +   b_0\sigma_1.
\end{equation*}
Here $m_0$ and $b_0$ are such real numbers:
\begin{equation*}
	\begin{aligned}
		m_0  = &  2 m_1 \langle \alpha_{\sharp}(\zeta), \partial_{\zeta}^2\alpha_{\sharp}(\zeta) \rangle_{L^2(\R)}  - 4\sgn(c_{\sharp})\Im(b_3)  \langle \alpha_{\sharp}(\zeta), \partial_{\zeta} \big( \eta(\zeta) \alpha_{\sharp}(\zeta) \big)\rangle_{L^2(\R)} \\  &  +2m_4 \langle \alpha_{\sharp}(\zeta),  \eta(\zeta)^2 \alpha_{\sharp}(\zeta) \rangle_{L^2(\R)} + 2\langle \alpha_{\sharp}(\zeta), \partial_{\zeta}^2 \alpha_{\sharp}(\zeta) \rangle_{L^2(\R)}  ;\\
		b_0  = & 2 \Re(b_1) \langle \alpha_{\sharp}(\zeta), \partial_{\zeta}^2\alpha_{\sharp}(\zeta) \rangle_{L^2(\R)}  .
	\end{aligned}
\end{equation*}

Thus, $\Det B_e =0$ has two real solutions $\mce^{(2)}_1(0) = -m_0+b_0$ and $\mce^{(2)}_1(0) = -m_0-b_0$.\quad\quad
  
\qed

{\bf (The second-order bifurcation of edge states' energies)} These two solutions are different if and only if:
\begin{equation}\label{eqn-main-1}
	\Re\bigg(\langle \phi_1(\bx), 2\bm{l}_2\cdot \nabla (\mch_V-E_{_D})^{-1} \mcq_{\perp}  2\bm{l}_2 \cdot \nabla \overline{\phi_1(-\bx)} \rangle _{L^2(\Omega)} \bigg)\neq 0 ,
\end{equation}
and 
\begin{equation}\label{eqn-main-2}
	\langle \alpha_{\sharp}(\zeta), \partial_{\zeta}^2\alpha_{\sharp}(\zeta) \rangle_{L^2(\R)} \neq 0 .
\end{equation}
Note that $\langle \alpha_{\sharp}(\zeta), \partial_{\zeta}^2\alpha_{\sharp}(\zeta) \rangle_{L^2(\R)} = 0$ if and only if $ \partial_{\zeta}\alpha_{\sharp}(\zeta)=0$ is true for almost all $\zeta \in \R $, which is certainly not true. Assume the condition (\ref{eqn-main-1}) is true throughout the following discussion so that the two edge states separate at order $O(\delta^2)$. Then we can solve $\bigl( c^1_1, c^2_1 \bigr)$ and $\bigl( c^1_2, c^2_2 \bigr)$ for equation (\ref{eqn-delta-2-3}), and $\bm{\beta}_1(\zeta;0)$ and $\bm{\beta}_2(\zeta;0)$ for equation (\ref{eqn-delta-2-2}) related to $\mce^{(2)}_1(0)$ and $\mce^{(2)}_2(0)$ respectively.

\subsection{Order $O(\delta^n)$ terms}

For edge states with separated energies, repeat using ansatz similar to (\ref{eqn-ansatz-1}) and the solvable conditions for order $O(\delta^n)$ can solve the multiscale problem recursively. The critical fact is that there are two independent eigenvalue problems with eigenvalues differing from at least at order $O(\delta^2)$.

\section{Rigorous formulation of two gapped edge states} \label{rigorous}

Based on the preparation in the last two sections, we are now ready to establish the existence of two gapped edge states at $k_{\parallel} = 0$ rigorously. Recall that $\chi_e$ is the function space at $k_{\parallel}=0$; see (\ref{eqn-chi-e}). The main theorem stating the existence of two edge states in $\chi_e$ is below. 

\begin{theorem}\label{thm-main}{\bf (Existence of the gapped edge states)}
    Let $H_V = -\Delta + V(\bx)$ be a bulk Hamiltonian as in Theorem \ref{thm-second-order-cone}, $H^{\delta}= -\Delta + V(\bx) +\delta W(\bx) $ be a bulk Hamiltonian as in Theorem \ref{thm-perturbation}, and the folding symmetry breaking domain wall modulated edge operator $\mch_{edge}^{\delta}=-\Delta + V(\bx) + \delta \eta(\delta\bm{l}_2 \cdot \bx) W(\bx)$ be as in Definition \ref{def-edge-operator}. Suppose that when $\phi_1(\bx)$ is chosen to satisfy (\ref{eqn-vSharp-l2}), the corresponding second-order non-degeneracy condition
    \begin{equation}\label{eqn-main}
	\Re\bigg(\langle \phi_1(\bx), 2\bm{l}_2\cdot \nabla (\mch_V-E_{_D})^{-1} \mcq_{\perp}  2\bm{l}_2 \cdot \nabla \overline{\phi_1(-\bx)} \rangle _{L^2(\Omega)} \bigg)\neq 0 
	\end{equation}
    is true.
    Then there exists $\delta_0>0$, such that for all $0< \delta < \delta_0$, $\mch_{edge}^{\delta}$ has two eigen pairs $\bigl(\mce_1,\psi_1(\bx)\bigr)$ and $\bigl(\mce_2,\psi_2(\bx)\bigr)$ in $\chi_e$  satisfying:
    \begin{enumerate}
        \item the energies $\mce_1$ and $\mce_2$ are near $E_D$ and gapped of order $O(\delta^2)$:
        \begin{equation}
            \mce_1 = E_{_D} + \mce_1^{(2)}\delta^2 + o(\delta^2), \quad \mce_2 = E_{_D} + \mce_2^{(2)}\delta^2 + o(\delta^2),
        \end{equation}
        where $\mce_1^{(2)}\neq \mce_2^{(2)}$;
        \item $\psi_1(\bx)$ and $\psi_2(\bx)$ are well-approximated by two slow modulations of linear combinations of a basis of  $\Ker(\mch_V -E_{_D})$:
        \begin{equation}
            \begin{aligned}
                \psi_1(\bx) & = c^1_1 \sum_{j=1}^4 \bm{\alpha}^1_j(\zeta) \phi_j(\bx)   + c^2_1\sum_{j=1}^4\bm{\alpha}^2_j(\zeta) \phi_j(\bx)   + O_{H^2(\R^2/{\Z \bm{w}_1})}(\delta^{\frac{1}{2}}),\\
                \psi_2(\bx) & = c^2_1\sum_{j=1}^4 \bm{\alpha}^1_j(\zeta)\phi_j(\bx)  + c^2_2\sum_{j=1}^4 \bm{\alpha}^2_j(\zeta) \phi_j(\bx)  + O_{H^2(\R^2/{\Z \bm{w}_1})}(\delta^{\frac{1}{2}}),
               \end{aligned}
        \end{equation}
       where $\{\phi_j(\bx)\}$ are the basis of $\Ker(\mch_V-E_{_D})$ as in the Theorem \ref{thm-doublecone}, $\zeta = \delta \bm{l}_2\cdot\bx$ is the slow variable, and $\bm{\alpha}^l(\zeta)$ are two orthonormal topologically protected zero-energy eigenstates of Dirac operator $\mcd(0)$ in (\ref{eqn-alpha}).
    \end{enumerate}
\end{theorem}

\begin{remark}
	The second order non-degeneracy condition (\ref{eqn-main}) guarantees $\mce_1^{(2)}\neq \mce_2^{(2)}$. The form of this condition changes with the phase transformation of $\phi_1(\bx)$, and therefore we fix a $\phi_1(\bx)$ which makes (\ref{eqn-vSharp-l2}) valid. 
 \end{remark}

\begin{figure}[htbp]
    \centering
    \subfigure[]{\includegraphics[width=5.3cm]{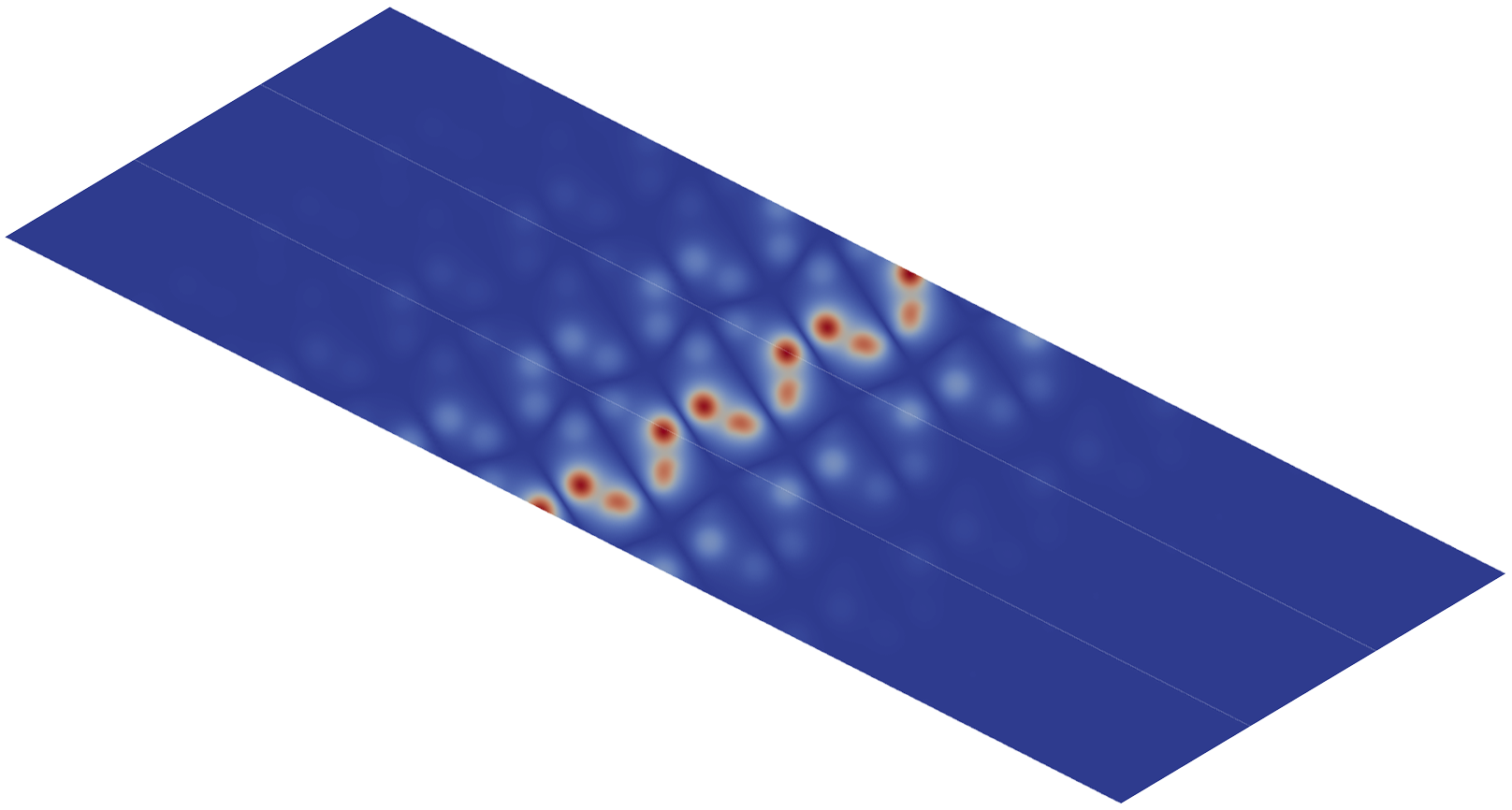}}\qquad\quad
    \subfigure[]{\includegraphics[width =5.3cm]{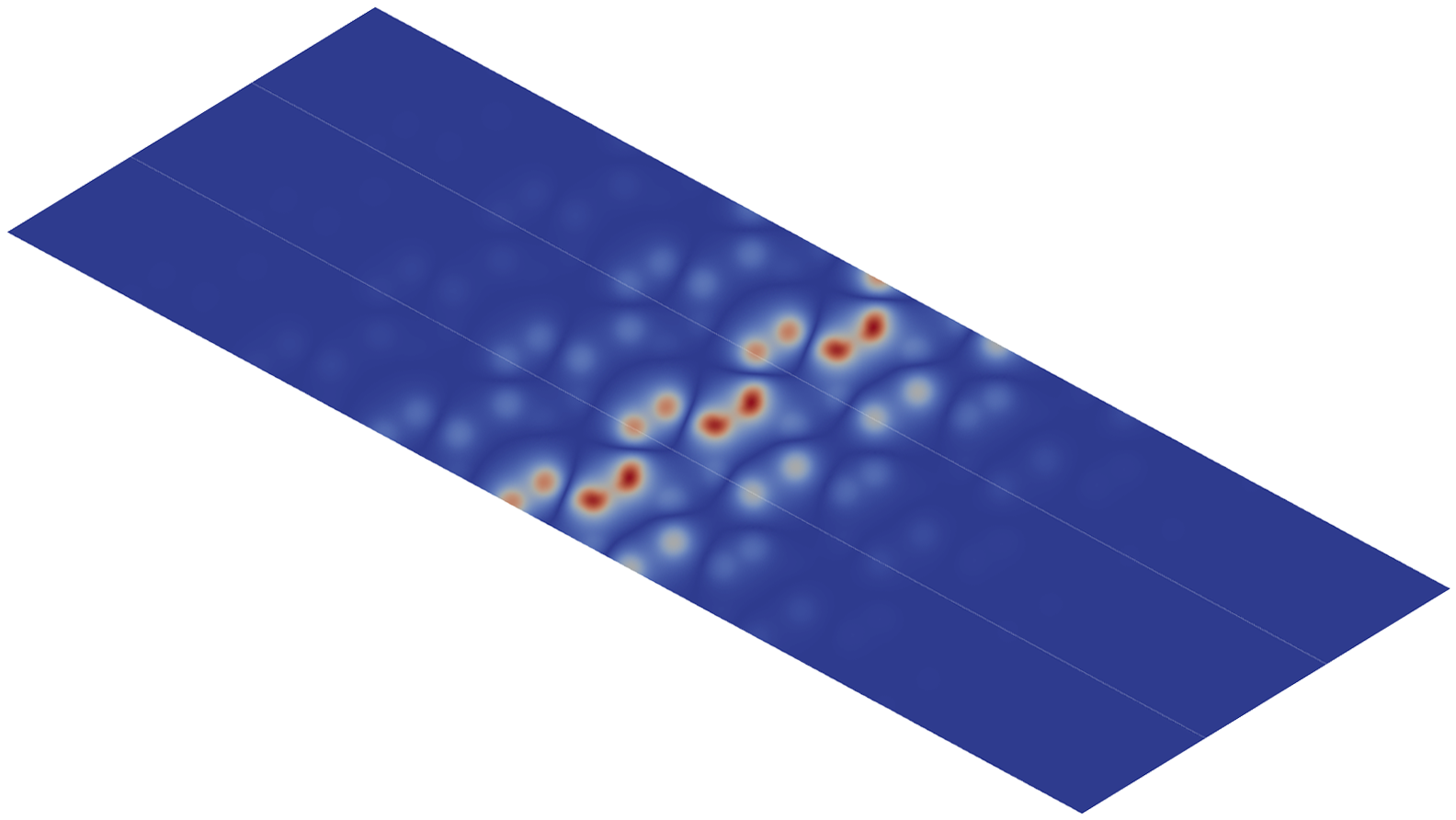}}\\
    \subfigure[]{\includegraphics[width =5.3cm]{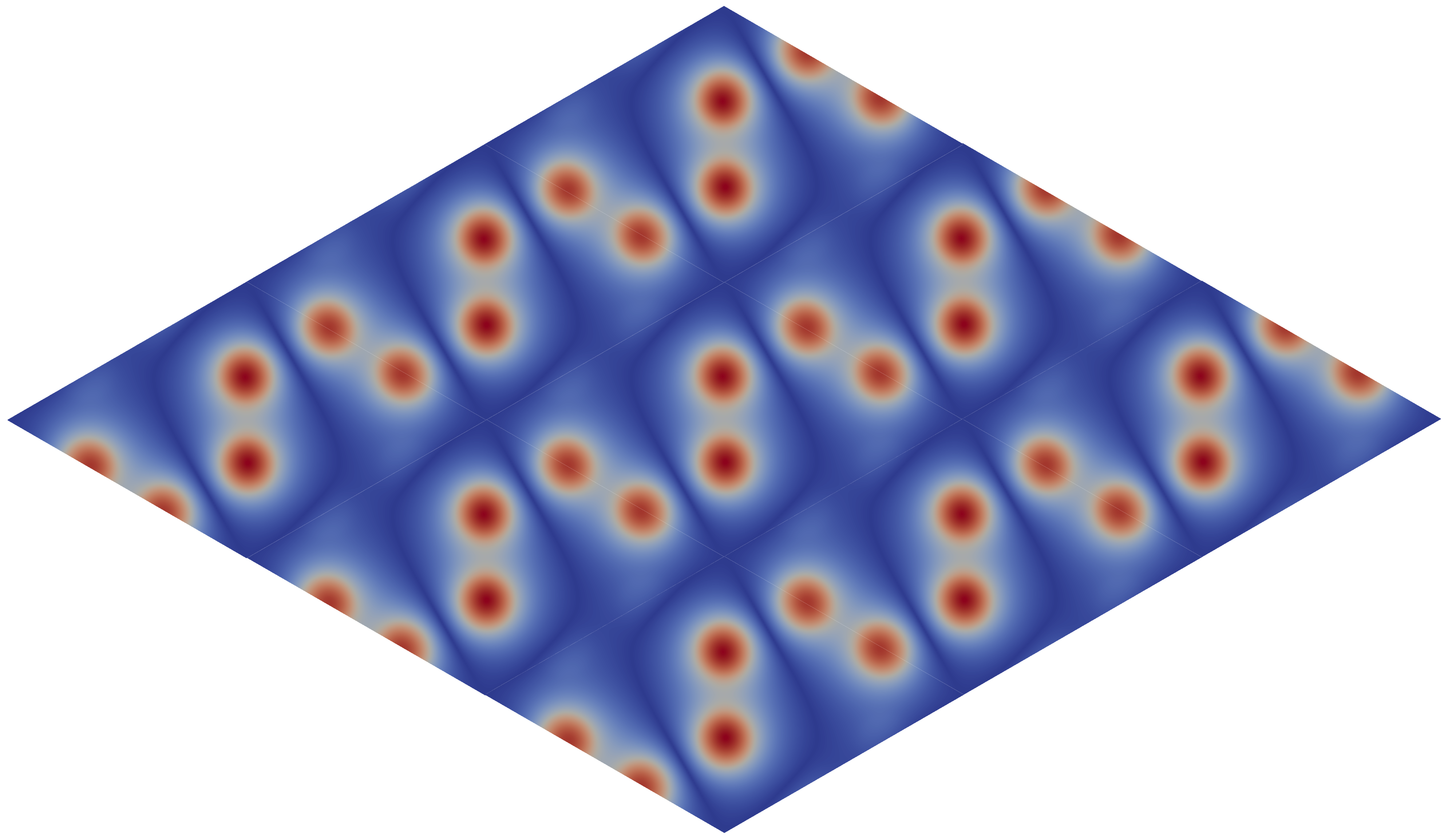}}\qquad\quad
    \subfigure[]{\includegraphics[width =5.3cm]{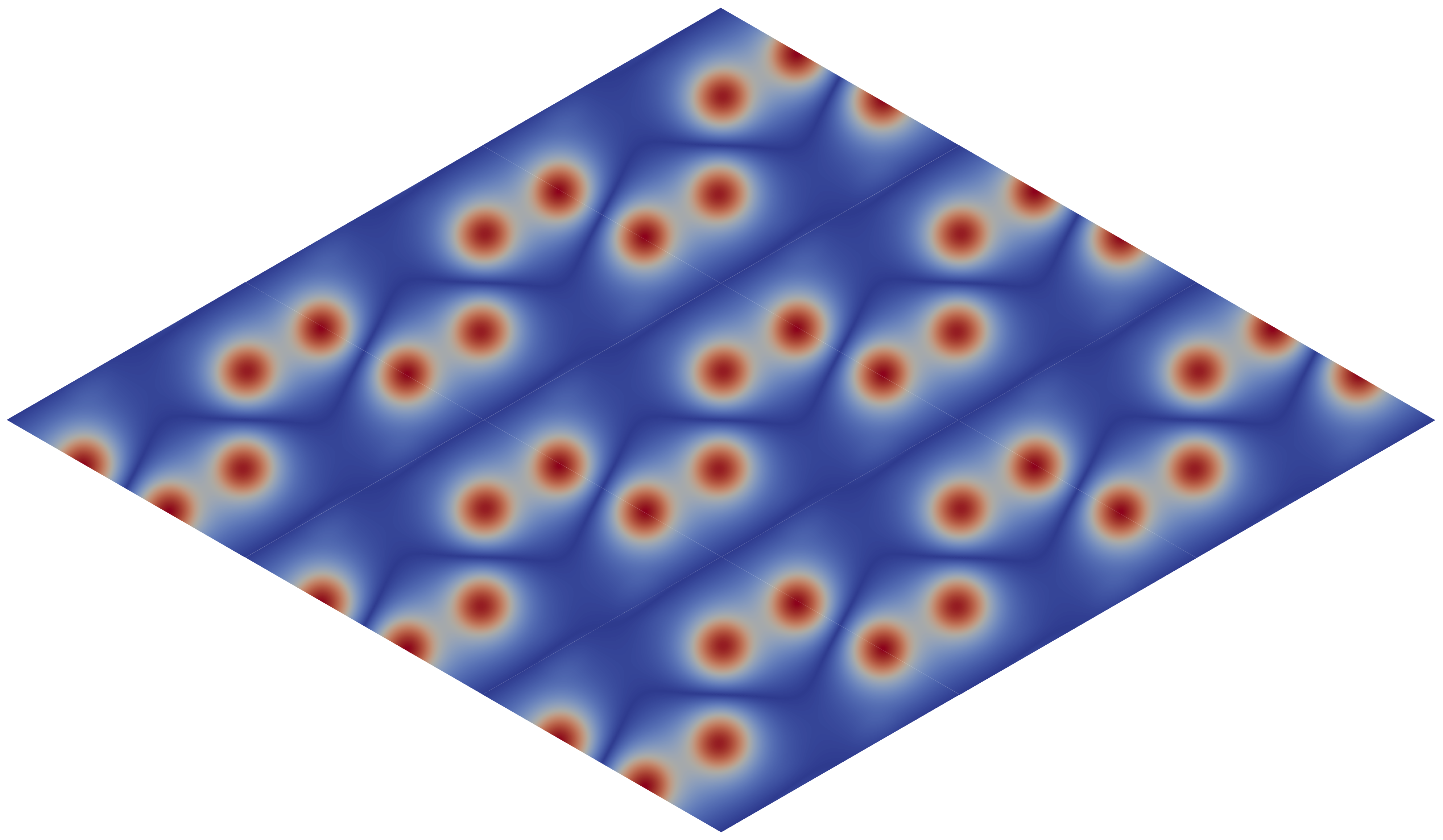}}\\
    \subfigure[]{\includegraphics[width =5.3cm]{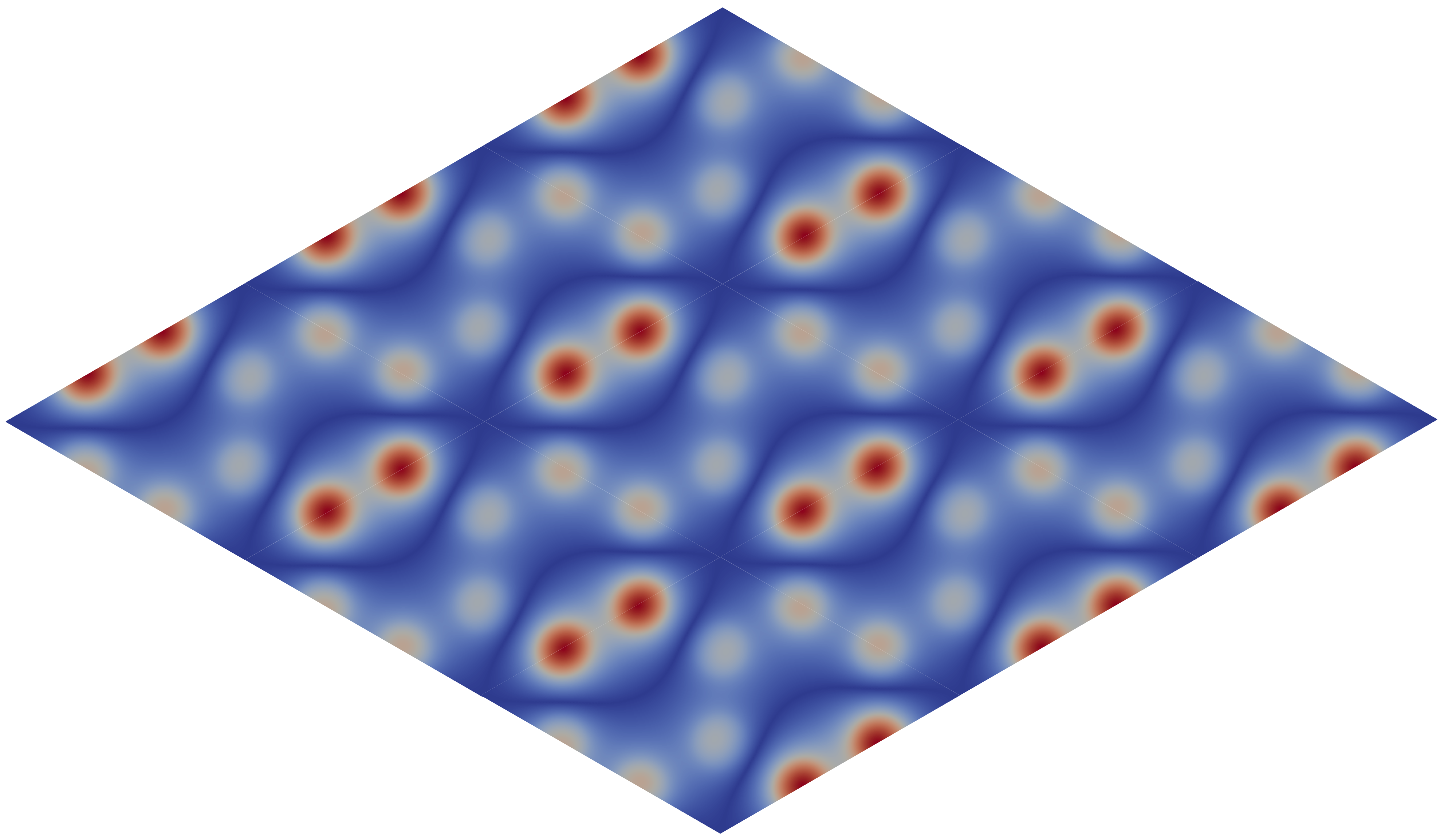}}\qquad\quad
    \subfigure[]{\includegraphics[width =5.3cm]{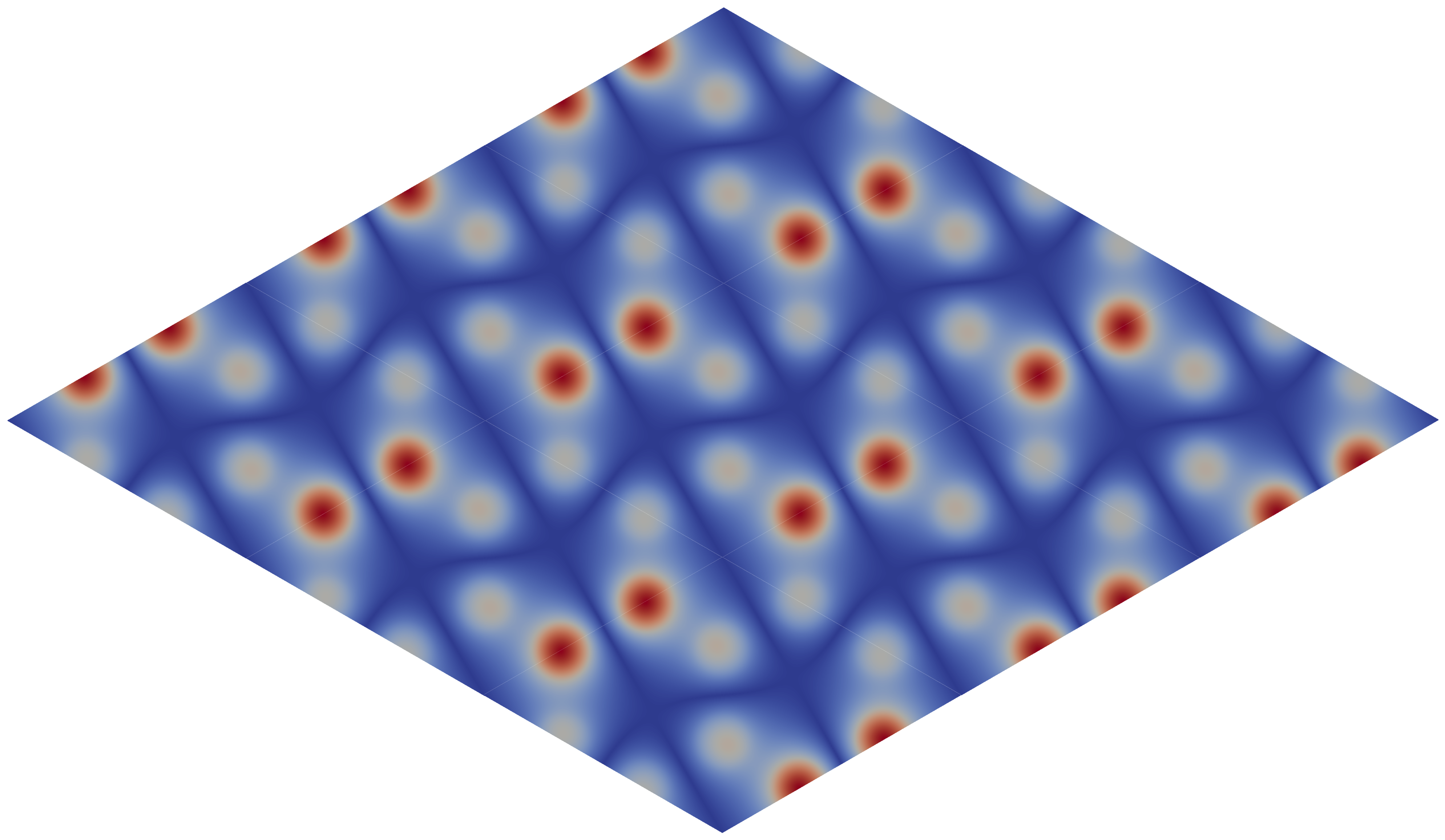}}
    \caption{Figures of edge and bulk modes. (a) and (b) the figures of numerical solutions of two edge states at $k_{\parallel} =0$ of the limiting domain wall model in Figure \ref{figure-energy-curve}. (c), (d), (e), and (f) the figures of numerical solutions of fourfold degenerate bulk modes of corresponding unperturbed bulk operators. The eigenstates in figure (a) and figure (b) are modulations of the eigenstates in figure (c) and figure (d), respectively.}
    \label{figure-ev}
\end{figure}

 This theorem states that $\mch_{edge}^{\delta}$ has two gapped edge states at $k_{\parallel}=0$ when $\delta$ is sufficiently small and characterizes the two edge modes by degenerate eigenmodes of $\mch_V$. Before giving the comprehensive proof, we show some numerical results in Figure \ref{figure-ev} that explicitly illustrate the conclusions. Figure (a) and (b) are the two gapped edge states at $k_{\parallel}=0$ near the edge. Figure (c)-(f) are a basis of the four-dimensional eigenspaces of the unperturbed bulk operator with eigenvalue $E_{_D}$. We choose a particular basis of the four-dimensional eigenspaces such that the eigenstates in figure (a) and figure (b) are modulations of the eigenstates in figure (c) and figure (d) respectively.
  
From the last section, we can obtain formal expansions of the two edge states at $k_{\parallel}=0$:
\begin{equation}\label{eqn-ri-ev1}
        \mce_1(\delta) = E_{_D} + \delta^2 \mu_1, \quad
        \psi_1(\bx,\zeta; \delta) = \psi^{(0)}_1(\bx, \zeta) + \delta \psi^{(1)}_1(\bx, \zeta) + \delta g(\bx);    
\end{equation}
\begin{equation}\label{eqn-ri-ev2}
    \mce_2(\delta) = E_{_D} + \delta^2 \mu_2, \quad
    \psi_2(\bx,\zeta; \delta) = \psi^{(0)}_2(\bx, \zeta) + \delta \psi^{(1)}_2(\bx, \zeta) + \delta h(\bx);
\end{equation}
where the $O(1)$ and $O(\delta)$ terms are as in the last subsection:
\begin{equation}\label{eqn-ri-ansazt1}
   \begin{aligned}
    \psi^{(0)}_1(\bx, \zeta) & = c^1_1 {\Phi}(\bx) ^{\mathrm{T}} \bm{\alpha}^1(\zeta) + c^2_1 {\Phi}(\bx) ^{\mathrm{T}} \bm{\alpha}^2(\zeta),\\
    \psi^{(1)}_1(\bx, \zeta) & = \Phi(\bx)  ^{\mathrm{T}} \bm{\beta}_1(\zeta;0)+ (\mch_V-E_{_D})^{-1}\mcl_1(s)\psi^{(0)}_1(\bx,\zeta);
   \end{aligned}
\end{equation}
\begin{equation}\label{eqn-ri-ansazt2}
    \begin{aligned}
        \psi^{(0)}_2(\bx, \zeta) & = c^1_2 {\Phi}(\bx) ^{\mathrm{T}} \bm{\alpha}^1(\zeta) + c^2_2 {\Phi}(\bx) ^{\mathrm{T}} \bm{\alpha}^2(\zeta),\\
        \psi^{(1)}_2(\bx, \zeta) & = \Phi(\bx)  ^{\mathrm{T}} \bm{\beta}_2(\zeta;0)+ (\mch_V-E_{_D})^{-1}\mcl_1(s)\psi^{(0)}_2(\bx,\zeta).
    \end{aligned}
\end{equation}

Substituting (\ref{eqn-ri-ev1}) into the eigenvalue problem at $k_{\parallel}=0$, we can obtain the corrector equation:
\begin{equation}\label{eqn-corrector}
    \begin{aligned}
        & (\mch_V -E_{_D})g(\bx) + \delta \eta(\delta \bm{l}_2 \cdot \bx) W(\bx)g(\bx) - \delta^2 \mu_1 g(\bx)\\
         = & \delta\bigl(\Vert \bm{l}_2\Vert^2 \partial_{\zeta}^2 + \mu_1\bigr) \psi^{(0)}(\bx,\zeta)|_{\zeta= \delta \bm{l}_2 \cdot \bx} +  \delta^2 \bigl(\Vert \bm{l}_2\Vert^2 \partial_{\zeta}^2 + \mu_1\bigr) \psi^{(1)}(\bx,\zeta)|_{\zeta= \delta \bm{l}_2 \cdot \bx}  \\ & + \delta\bigl(2 \bm{l}_2 \cdot \nabla_{\bx} \partial_{\zeta}  - \eta(\zeta)W(\bx) \bigr)\psi^{(1)}(\bx,\zeta)|_{\zeta= \delta \bm{l}_2 \cdot \bx}\quad ,
    \end{aligned}
\end{equation} 
and the same equation for $\mu_2$ and $h(\bx)$. It remains to solve this equation and estimate the order of $g(\bx)$. We only need to construct rigorous results for $(\ref{eqn-corrector})$, and the same can be done for $\mu_2$ and $h(\bx)$. The idea is to decompose $g(\bx)$ by Floquet-Bloch modes and decompose the equation into different components accordingly \cite{Fefferman2016}. Far-energy components can be solved as a functional of near-energy components. See section \ref{near-energy} for the description of near-energy approximation. Finally, we get a closed system of near-energy components, where we can use Lyapunov-Schmidt reduction. 

It is obvious that $g(\bx)$ should be in $\chi_e$. The following lemma shows that Floquet-Bloch eigenmodes are a complete basis for $\chi_e$. According to this lemma, equation (\ref{eqn-corrector}) can be decomposed into a family of equations.

\begin{lemma}
    For $f\in \chi_e$, where $\chi_e$ is defined in (\ref{eqn-chi-e}), the following decomposition is true:
    \begin{equation}
        f(\bx) =\sum_{n\geq 1} \int_{-\frac{1}{2}}^{\frac{1}{2}}  \Tilde{f}_n(\lambda) e_n(\bx; \lambda \bm{l}_2) d \lambda.
    \end{equation}
Here $\big\{ e_n(\bx; \bk)\big\}_{n\in \N^*}$ are a complete orthonormal basis of $L^2_{\bk}(\R^2/ {\bf U})$ given by normarlized eigenstates of $\mch_V$ on $L^2_{\bk}(\R^2/ {\bf U})$. They are called Floquet-Bloch modes. And
\begin{equation}
    \Tilde{f}_n(\lambda) = \langle f(\bx), e_n(\bx;\lambda \bm{l}_2) \rangle_{L^2({\Omega_e})}.
\end{equation}
\end{lemma}

For eigenmodes $e_n(\bx;\lambda\bm{l}_2)$ with eigenvalue near $E_{_D}$, there exists near energy approximations; see Proposition \ref{prop-near-appro}. We rearrange the spectrum and complete orthonormal eigenmodes accordingly: 
\begin{equation*}
    E_n(\lambda) = \begin{cases}
        \theta_j(\lambda),\quad n=n_*+j,\quad j=1,2,3,4, \quad |\lambda | \leq \delta^{\nu}\\
        E_n(\lambda \bm{l}_2),\quad else;
    \end{cases}
\end{equation*}
\begin{equation*}
    e_n(\bx;\lambda) = \begin{cases}
        \Theta_j(\bx;\lambda),\quad n=n_*+j, \quad j=1,2,3,4, \quad |\lambda | \leq \delta^{\nu}\\
        e_n(\bx; \lambda \bm{l}_2),\quad else.
    \end{cases}
\end{equation*}

Now let us take inner products of (\ref{eqn-corrector}) with $e_n(\bx; \lambda)$ to obtain equations of $\{\tilde{g}_n(\lambda) \}_n$:
\begin{equation}\label{eqn-co-all}
    \begin{aligned}
    &(E_n(\lambda) - E_{_D}) \tilde{g}_n(\lambda) + \delta \langle e_n(\cdot;\lambda ), \eta(\delta \bm{l}_2 \cdot )W(\cdot)g(\cdot) \rangle_{L^2(\Omega_e)} \\
    = & \delta F_n(\lambda;\delta,\mu_1) + \delta^2 \mu_1\tilde{g}_n(\lambda).
    \end{aligned}
\end{equation}
Here
\begin{equation*}
    \begin{aligned}
        F_n(\lambda;\delta,\mu_1)  & = F_n^1(\lambda; \delta, \mu_1 ) +   F_n^2(\lambda; \delta, \mu_1 ) + \delta F_n^3(\lambda; \delta, \mu_1 ),\\
        F_n^1(\lambda; \delta, \mu_1 ) & = \langle e_n(\bx;\lambda),  \bigl(\Vert \bm{l}_2\Vert^2 \partial_{\zeta}^2 + \mu_1\bigr) \psi^{(0)}(\bx,\zeta)|_{\zeta= \delta \bm{l}_2 \cdot \bx}  \rangle_{L^2(\Omega_e)},\\
        F_n^2(\lambda; \delta, \mu_1 ) &= \langle e_n(\bx;\lambda ),  \bigl(2 \bm{l}_2 \cdot \nabla_{\bx} \partial_{\zeta}  - \eta(\zeta)W(\bx) \bigr)\psi^{(1)}(\bx,\zeta)|_{\zeta= \delta \bm{l}_2 \cdot \bx} \rangle_{L^2(\Omega_e)},\\
        F_n^3(\lambda; \delta, \mu_1 ) &= \langle e_n(\bx;\lambda ),  \bigl(\Vert \bm{l}_2\Vert^2 \partial_{\zeta}^2 + \mu_1\bigr) \psi^{(1)}(\bx,\zeta)|_{\zeta= \delta \bm{l}_2 \cdot \bx}   \rangle_{L^2(\Omega_e)}.
    \end{aligned}
\end{equation*}
$\Tilde{g}_n(\lambda)$ can be decomposed into several parts $\Tilde{g}_n(\lambda) = \sum_{j=1}^4\Tilde{g}_{j,near}(\lambda) + \Tilde{g}_{n,far}(\lambda)  $:
\begin{equation}
    \Tilde{g}_{j,near}(\lambda) =\chi_{near}(\delta) \Tilde{g}_{n_*+j}(\lambda),
\end{equation}
\begin{equation}
    \Tilde{g}_{n,far}(\lambda) = \chi_{n,far}(\delta)  \Tilde{g}_n(\lambda).
\end{equation}
Here $$\chi_{near}(\delta)  =  \chi(|\lambda| \leq \delta^{\nu}) ,\quad \chi_{n,far} (\delta) = \chi\biggl( \bigl(\sum_{j=1}^4\delta_{n,n_*+j} \bigr) \delta^{\nu}\leq \lambda \leq \frac{1}{2} \biggr), $$ $\nu$ is chosen appropriately by spectral no-fold condition (\ref{eqn-non-fold}) \cite{Drouot2020,Fefferman2016} and $\delta_{n,n_*+j}$ are Kronecker delta symbols. Accordingly we obtain $g(\bx) = g_{near}(\bx) + g_{far}(\bx)$:
\begin{equation}\label{eqn-g-near}
    g_{near}(\bx) = \sum_{j=1}^4 \int_{-\frac{1}{2}}^{\frac{1}{2}} \Tilde{g}_{j,near}(\lambda) \Theta_{j}(\bx;\lambda )  d \lambda ,
\end{equation}
\begin{equation}\label{eqn-g-far}
    g_{far}(\bx) = \sum_{n\geq 1}\int_{-\frac{1}{2}}^{\frac{1}{2}} \Tilde{g}_{n,far}(\lambda) e_{n}(\bx;\lambda )  d \lambda.
\end{equation}

Therefore, we can divide equations (\ref{eqn-co-all}) into near energy components:
\begin{equation}\label{eqn-co-near}
    \begin{aligned}
        (\theta_{j}(\lambda & )  - E_{_D}) \tilde{g}_{j,near}(\lambda) \\
        + & \delta \chi_{near}(\delta) \langle \Theta_{j}(\cdot;\lambda ), \eta(\delta \bm{l}_2 \cdot )W(\cdot)\bigl( g_{near}(\cdot) + g_{far}(\cdot) \bigr) \rangle_{L^2(\Omega_e)} \\
        =  \chi_{near}& (\delta)  \delta  F_{n_*+j}(\lambda;\delta,\mu_1)
        +  \delta^2 \mu_1\tilde{g}_{j,near} (\lambda);
        \end{aligned}
\end{equation}
and far energy components:
\begin{equation}\label{eqn-co-far}
    \begin{aligned}
        (E_n(\lambda & )  - E_{_D}) \tilde{g}_{n,far}(\lambda) \\
        + & \delta \chi_{n,far}(\delta) \langle e_n(\cdot;\lambda ), \eta(\delta \bm{l}_2 \cdot )W(\cdot)\bigl( g_{near}(\cdot) + g_{far}(\cdot) \bigr) \rangle_{L^2(\Omega_e)} \\
        =  \chi_{n,far}& ( \delta )  \delta  F_n(\lambda;\delta,\mu_1)
        +  \delta^2 \mu_1\tilde{g}_{n,far} (\lambda) .
        \end{aligned}
\end{equation}

Rewrite the equation of far-energy components (\ref{eqn-co-far}) as:
\begin{equation*}
    \begin{aligned}
        -  \delta \frac{  \chi_{n,far}(\delta)}{E_n(\lambda )  - E_{_D}}\langle e_n(\cdot;\lambda ), \eta(\delta \bm{l}_2 \cdot )W(\cdot)\bigl( g_{near}(\bx) + g_{far}(\cdot) \bigr) \rangle_{L^2(\Omega_e)}& \\
          + \delta \frac{ \chi_{n,far}(\delta )   }{E_n(\lambda )  - E_{_D}}   F_n(\lambda;\delta,\mu_1)   +  \delta^2 \frac{\mu_1\tilde{g}_{n,far} (\lambda)}{E_n(\lambda )  - E_{_D}}
            & =  \tilde{g}_{n,far} (\lambda) .
    \end{aligned} 
\end{equation*}
This can be viewed as a fixed problem of $\{\Tilde{g}_{n,far}(\lambda)\}$ and easily transformed into a fixed point problem of $g_{far}(\bx)$ by (\ref{eqn-g-far}). Fixing $\delta$ and $\mu_1$, via contraction mapping principle, $g_{far}(\bx)$ can be solved out as a functional of $g_{near}(\bx)$: $g_{far}(\bx) = g_{far}[g_{near}; \mu_1,\delta](\bx)$. 

Now let us look at (\ref{eqn-co-near}) -- the equation of near-energy components. Substituting $g_{far}(\bx) = g_{far}[g_{near}; \mu_1,\delta](\bx)$ into (\ref{eqn-co-near}), using the rescaling $\xi = \frac{\lambda}{\delta}$ and the results in Proposition \ref{prop-near-appro} and canceling a factor of $\delta$, we can finally obtain the closed system of near-energy components: 
\begin{equation}
    \begin{aligned}
    & -v_{_F} \Vert \bm{l}_2 \Vert \xi \Tilde{g}_{j,near}(\delta \xi)  + \chi_{near}(\delta)\langle \Theta_j(\cdot;\delta\xi), \eta(\bm{l}_2\cdot)W(\cdot) g_{near}(\cdot) \rangle_{L^2(\Omega_e)} \\
    = & \chi_{near}(\delta) F_{b+j}(\delta\xi; \delta,\mu_1) - \delta r_j(\delta\xi)\xi^2 \Tilde{g}_{j,near}(\lambda) + \delta \mu_1 \Tilde{g}_{j,near}(\delta\xi)\\
    & - \chi_{near}(\delta)\langle \Theta_j(\cdot;\delta\xi), \eta(\bm{l}_2\cdot)W(\cdot) g_{far}[g_{near};\mu_1,\delta](\cdot) \rangle_{L^2(\Omega_e)}.
    \end{aligned}
\end{equation}
The rest of the steps are transforming this system into a Dirac system, solving the system by Lyapunov-Schmidt reduction, and obtaining upper bound estimations of $\mu_1$ and $g(\bx)$, which is similar to the $\mcp\mct$ symmetry breaking case \cite{Fefferman2016}.

{
\section{Topological interpretations}\label{physical-intepretation}
In this section, we try to interpret our analytical results obtained in precious sections from topological perspectives.  Such an interpretation may provide deeper insights into the intrinsic properties of the two edge states. We include two aspects: one is the topological indices-a traditional approach that may reveal bulk-edge correspondence; the other is the parities of the eigenmodes - which can be clarified only in special symmetry breaking cases, such as the folding symmetry breaking. These two aspects exhibit a high degree of consistency with each other.

\subsection{The effective local topological charge}\label{sec-index}
Building on the work of \cite{Drouot2019}, the effective model can be used to approximate the local contribution to the Chern number of eigenfunction fiber bundle of $\mch^{\delta}$ over the Brillouin zone. This Chern number is exactly the topological index of the bulk operator. And let us call the local contribution approximated by effective models at the $\Gamma$ point here as an effective local topological charge.	In this subsection, we will establish the local charge step by step and demonstrate the critical importance of the second-order approximation.

Let us begin with the dispersion eigenvalue problem of perturbed bulk operator $\mch^{\delta} = -\Delta + V(\bx) + \delta W(\bx)$ on $L_{\bk}^2(\R^2 / {\bf U})$:
\begin{equation*}
	\begin{aligned}
		\mch^{\delta} \phi(\bx;\bk) &= E(\bk)\phi(\bx;\bk),\\
		\phi(\bx;\bk) = e&^{i\bk\cdot\bx} p(\bx),\quad p(\bx)  \in \chi,
	\end{aligned}
\end{equation*}
which is equivalent to 
\begin{equation*}
		\mch^{\delta}(\bk) p(\bx;\bk) = E(\bk) p(\bx;\bk),\quad p(\bx) \in \chi,
\end{equation*}
where $\mch^{\delta}(\bk) = -(\nabla + i\bk)^2 + V(\bx)+ \delta W(\bx)$.

We only need to consider the near-energy solution:
\begin{equation*}
	E(\bk) = E_{_D} + \mu, \quad p(\bx;\bk) =   {\Phi}(\bx) ^{\mathrm{T}}\bf{P}(\bk) + \psi(\bx;\bk).
\end{equation*}
By reduction procedures similar to those in Section \ref{sec-2nd-birfurcation}, we can obtain the equivalent equations:
$$(\mu I + B^{\delta}(\bk)) {\bf P}(\bk) = {\bf 0}$$
with $B^{\delta}(\bk) = B_1^{\delta}(\bk) + B_2^{\delta}(\bk) + O(\Vert \bk \Vert^3)$, where the first and second-order bifurcation matrices are as below:
\begin{equation*}\label{eqn-bifur-delta}
	B^{\delta}_1(\bk) = \begin{pmatrix}
	0 & 2i \bk\cdot \bm{v}_{\sharp} & -\delta c_{\sharp} & 0 \\
	\overline{2i\bk\cdot  \bm{v}_{\sharp}} & 0 & 0 & -\delta c_{\sharp}\\
	-\delta c_{\sharp} & 0 & 0 & -2i\bk\cdot  \bm{v}_{\sharp}\\
	0 & -\delta c_{\sharp} & \overline{-2i\bk\cdot  \bm{v}_{\sharp}} & 0
	\end{pmatrix};
\end{equation*}
\begin{equation*}
	B^{\delta}_2(\bk)= (m(\bk) -\Vert \bk \Vert^2 )I + \begin{pmatrix}
			0 & b(\bk) & 0 & 0\\
			\overline{b(\bk)} & 0 & 0 & 0 \\
			0 & 0 & 0& b(\bk) \\
			 0 &  0 &  \overline{b(\bk)} & 0
		\end{pmatrix}.
\end{equation*}
Here $m(\bk)$ and $b(\bk)$ are the second-order term as in Section \ref{sec-2nd-birfurcation}. These matrices can be transformed by $Q$ mentioned in (\ref{eqn-trans-Q}) into:
\begin{equation}\label{eqn-b-tlide-1}
	\Tilde{B}^{\delta}_1(\bk) = Q^{T} B^{\delta}_1(\bk)Q = \begin{pmatrix}
		-\delta c_{\sharp} & 2i\bk\cdot\bm{v}_{\sharp} & 0 & 0\\
		\overline{2i\bk\cdot \bm{v}_{\sharp}} & \delta c_{\sharp} & 0 & 0\\
		0 & 0& \delta c_{\sharp} & 2i\bk\cdot\bm{v}_{\sharp}\\
		0 & 0& \overline{2i\bk\cdot\bm{v}_{\sharp}} & -\delta c_{\sharp}
	\end{pmatrix} ;
\end{equation}
\begin{equation}\label{eqn-b-tlide-2}
	\begin{aligned}
		\Tilde{B}_2^{\delta}(\bk) = Q^TB_2^{\delta}(\bk)Q =  (m(\bk) -\Vert \bk \Vert^2 )I + \begin{pmatrix}
			0 & 0 & 0 & b(\bk)\\
			0 & 0& \overline{b(\bk)} & 0\\
			0 & b(\bk) & 0 & 0\\
			\overline{b(\bk)} & 0 & 0& 0
		\end{pmatrix}.
	\end{aligned}
\end{equation}

The effective local topological charge of this problem is the following integral:
\begin{equation}\label{eqn-local-charge}
    \Pi_j =  \frac{1}{2\pi i}\lim_{r\to 0} \oint _{\partial D(0,r)}\langle \xi_j(\bk), \nabla_{\bk}  \xi_j(\bk) \rangle \cdot d \bk.
\end{equation}
Here $\xi_j(\bk)$ is the $j^{th}$ eigenvector of $\bigl(\Tilde{B}^{\delta}_1(\bk) + \Tilde{B}^{\delta}_2(\bk)\bigr)$, which needs to be normalized and have singularity at $\bk = {\bf 0}$. Our conclusion is as follows.

\begin{theorem}{\bf (Effective local topological charges)}
    For $\bigl(\Tilde{B}_1^{\delta}(\bk) + \Tilde{B}_2^{\delta}(\bk)\bigr)$ in (\ref{eqn-b-tlide-1}) and (\ref{eqn-b-tlide-2}), the local topological charges in (\ref{eqn-local-charge}) are:
    \begin{equation}
        \Pi_1 = \Pi_2 = -\sgn(\delta c_{\sharp}), \q \Pi_3 = \Pi_4 = \sgn(\delta c_{\sharp}).
    \end{equation}
\end{theorem}

This theorem states that the effective local topological charges of the upper pair bands and the lower pair bands are always identical within each pair and opposite to each other. We will prove this theorem by first considering $ \Tilde{B}^{\delta}_1(\bk) $ individually, and then considering $\Tilde{B}^{\delta}_1(\bk)$ and $\Tilde{B}^{\delta}_2(\bk)$ together to analyze the influence of second-order terms on the topology.

\subsubsection{First-order approximation}\label{sec-index-1}
We have seen that the first-order bifurcation matrix can be 
diagonalized into $
  \Tilde{B}^{\delta}_1(\bk)=\diag(B_s(\bk;\delta),B_s(\bk;-\delta))
$; where
\begin{equation}
    B_s(\bk;\delta) = \begin{pmatrix}
	-\delta c_{\sharp} & 2i\bk\cdot\bm{v}_{\sharp}\\
	\overline{2i\bk\cdot \bm{v}_{\sharp}} & \delta c_{\sharp}
\end{pmatrix}.
\end{equation}
From the diagonal elements, $\Tilde{B}_1^{\delta}(\bk)$ resembles a superposition of two perturbed Dirac operators. However, it is important to note that, despite their appearance of decoupling, they are actually coupled due to sharing the same eigenvalues. This intrinsic coupling arises from the $\mcp\mct$ symmetry-preserving property. To characterize the topology of each energy band individually, we must separate the bands through certain methods, such as considering higher-order terms, which can also clarify the nature of their coupling.

But now let us just look at the diagonal part $B_s(\bk;\delta)$. It is the Fourier transformation of a Dirac operator. The solvable condition $\Det(\lambda I + B_s(\bk;\delta)) =0 $ has two solutions $\lambda_{\pm} (\bk;
\delta)= \pm \sqrt{4\Vert \bk\cdot  \bm{v}_{\sharp}\Vert^2+\delta^2c_{\sharp}^2 }$. Suppose the corresponding normalized eigenvectors are $\xi_{\pm}(\bk;\delta)$, which should have singularities at $\bk = {\bf 0} $. The corresponding topological index is
\begin{equation}
    {\Pi}_{\pm} = \frac{1}{2\pi i}\lim_{r\to 0} \oint _{\partial D(0,r)}\langle \xi_{\pm}(\bk;\delta), \nabla_{\bk}  \xi_{\pm}(\bk;\delta) \rangle \cdot d \bk = \pm \sgn(\delta c_{\sharp}).
\end{equation}
The calculation of such topological indices for the energy bands of Dirac operators is standard; see, for example \cite{Drouot2019}.

\subsubsection{Second-order approximation}\label{sec-index-2}

We have known that $\Det(\lambda I + \Tilde{B}_1^{\delta}(\bk)) =0 $ have two solutions both of multiplicity two:
$$\lambda_{\pm}(\bk;\delta) = \pm \sqrt{4\Vert \bk\cdot\bm{v}_{\sharp} \Vert^2 + \delta^2 c_{\sharp}^2}.$$

Take $\sgn(\delta c_{\sharp})>0$ and the corresponding upper two bands with eigenvalues near $\lambda_+$ as an example to calculate the effective local topological charge in detail.

For the two diagonal elements of $\Tilde{B}_1^{\delta}(\bk)$, the corresponding topological indices of band $\lambda_+$ are $\pm 1$ respectively. The corresponding two normalized eigenvectors with singularities at $\bk = {\bf 0}$ are:
\begin{equation*}
    \xi_1^+ =\frac{1}{\sqrt{2\lambda_+(\lambda_+ -\delta c_{\sharp})}} \begin{pmatrix}
		\overline{-2i\bk\cdot\bm{v}_{\sharp}}\\\delta c_{\sharp} - \lambda_+\\ 0\\0
	\end{pmatrix} , \q
    \xi_2 ^+= \frac{1}{\sqrt{2\lambda_+(\lambda_+ -\delta c_{\sharp})}}\begin{pmatrix}
		0\\0\\\delta c_{\sharp} - \lambda_+\\-2i\bk\cdot\bm{v}_{\sharp}
	\end{pmatrix} .
\end{equation*}

Now including the second-order bifurcation matrix, we can solve $\bigl(\lambda I + \Tilde{B}_1^{\delta}(\bk) + \Tilde{B}_2^{\delta}(\bk)\bigr) \xi = 0$ with $\lambda$ near $\lambda_+$. By Lyapunov-Schdimt reduction, the associated eigenvectors are:
\begin{equation*}
	\xi_3 = \frac{1}{\sqrt{2}}\xi_1 + \frac{a_+(\bk;\delta)}{\sqrt{2}|a_+(\bk;\delta)|}\xi_2+ O(\Vert\bk\Vert^2), \q \xi_4 =\frac{1}{\sqrt{2}}\xi_1  - \frac{a_+(\bk;\delta)}{\sqrt{2}|a_+(\bk;\delta)|}\xi_2 + O(\Vert\bk\Vert^2).
\end{equation*}
Here $a_+(\bk;\delta)$ is of the form:
\begin{equation}\label{eqn-a-index}
	a_+(\bk;\delta) =\vert2i\bk\cdot \bm{v}_{\sharp}\vert^2\overline{b(\bk)}+(\delta c_{\sharp}-\lambda_+)^2{b(\bk)}.
\end{equation}
Note that 
\begin{equation*}
    \begin{aligned}
         \Pi_3 & =  \frac{1}{2\pi i}\lim_{r\to 0} \oint _{\partial D(0,r)}\langle \xi_3(\bk;\delta), \nabla_{\bk}  \xi_3(\bk;\delta) \rangle d \bk \\
          & =  \frac{1}{4\pi i}\lim_{r\to 0} \oint _{\partial D(0,r)}\biggl(\langle \xi_1^+, \nabla_{\bk}  \xi_1^+ \rangle  +  \langle \frac{a_+(\bk;\delta)}{|a_+(\bk;\delta)|}\xi_2^+, \nabla_{\bk}  \biggl(\frac{a_+(\bk;\delta)}{|a_+(\bk;\delta)|}\xi_2^+\biggr)\rangle   \biggr)d \bk \\
          & =   \frac{1}{2\pi i}\lim_{r\to 0} \oint _{\partial D(0,r)}\langle \xi_4(\bk;\delta), \nabla_{\bk}  \xi_4(\bk;\delta) \rangle d \bk = \Pi_4.
    \end{aligned}
\end{equation*}
Besides,
\begin{equation*}
    \begin{aligned}
         \Pi_3 & =  \frac{1}{2\pi i}\lim_{r\to 0} \oint _{\partial D(0,r)}\langle \xi_3(\bk;\delta), \nabla_{\bk}  \xi_3(\bk;\delta) \rangle d \bk \\
          & =  \frac{1}{4\pi i}\lim_{r\to 0} \oint _{\partial D(0,r)}\biggl(\langle \xi_1^+, \nabla_{\bk}  \xi_1^+ \rangle  +  \langle \frac{a_+(\bk;\delta)}{|a_+(\bk;\delta)|}\xi_2^+, \nabla_{\bk}  \biggl(\frac{a_+(\bk;\delta)}{|a_+(\bk;\delta)|}\xi_2^+\biggr)\rangle   \biggr)d \bk \\
          & = \frac{1}{4\pi i}\lim_{r\to 0} \oint _{\partial D(0,r)}\biggl(\langle \xi_1^+, \nabla_{\bk}  \xi_1^+ \rangle  +  \langle \frac{a_+(\bk;\delta)}{|a_+(\bk;\delta)|}\xi_2^+, \frac{a_+(\bk;\delta)}{|a_+(\bk;\delta)|}\nabla_{\bk}  \xi_2^+\rangle  \\
          & \qq\qq \qq\qq\qq\qq +\langle \frac{a_+(\bk;\delta)}{|a_+(\bk;\delta)|}, \nabla_{\bk} \frac{a_+(\bk;\delta)}{|a_+(\bk;\delta)|}\rangle  \langle \xi_2^+,\xi_2^+ \rangle \biggr)d \bk   \\
           \end{aligned}
\end{equation*}
\begin{equation*}
    \begin{aligned}
          & = \frac{1}{4\pi i}\lim_{r\to 0} \oint _{\partial D(0,r)} \langle \frac{a_+(\bk;\delta)}{|a_+(\bk;\delta)|}, \nabla_{\bk} \frac{a_+(\bk;\delta)}{|a_+(\bk;\delta)|}\rangle d \bk\qq\qq\qq\qq\qq \\
           &= \frac{1}{4\pi i}\lim_{r\to 0} \oint _{\partial D(0,r)}  \frac{\Im\bigl(a_+(\bk;\delta) \nabla_{\bk} \overline{a_+(\bk;\delta)}\bigr)}{|a_+(\bk;\delta)|^2} d \bk\\
           & = \frac{1}{4\pi i}\lim_{r\to 0} \oint _{\partial D(0,r)}  \frac{\Im \nabla_{\bk} \overline{a_+(\bk;\delta)}}{\overline{a_+(\bk;\delta)}} d \bk .
    \end{aligned}
\end{equation*}
Let  $\bk =\begin{pmatrix}
    r\cos{\theta}\\
    r \sin{\theta}
\end{pmatrix} $. The local charge is:
\begin{equation}
\begin{aligned}
    \Pi_3 & = \frac{1}{4\pi i}\lim_{r\to 0} \oint _{\partial D(0,r)}  \frac{\Im \partial_{\theta} \overline{a_+(\bk(r,\theta);\delta)}}{\overline{a_+(\bk(r,\theta);\delta)}} d \theta \\
    & =  - \frac{1}{4\pi i}\lim_{r\to 0} \oint _{\partial D(0,r)}  \frac{\Im \partial_{\theta} {a_+(\bk(r,\theta);\delta)}}{{a_+(\bk(r,\theta);\delta)}} d \theta.
\end{aligned}
\end{equation}
From the detailed calculation, we know that the local effective topological charge is half of the winding number of $\overline{a_+(\bk;\delta)}$. Here, from the equation (\ref{eqn-a-index}), $a_+(\bk;\delta)$ is:
\begin{equation*}
	a_+(\bk;\delta) =\vert2i\bk\cdot \bm{v}_{\sharp}\vert^2\overline{b(\bk)}+(\delta c_{\sharp}-\lambda_+)^2{b(\bk)}.
\end{equation*}
For the form of $b(\bk)$ in Appendix \ref{appen}, we can choose a typical $e^{i\theta^*}\phi_1(\bx)$ such that $b_0 \in \R$. $v_{_F}$ and $c_{\sharp}$ stay invariant under such phase transformations. Thus, we have:
\begin{equation*}
\begin{aligned}
     \Re a_+(\bk;\delta) & = b_0 ( v_{_F}^2 r^2 + (\delta c_{\sharp} - \lambda_+)^2 ) r^2 \sin{2\theta}, \\
     \Im a_+(\bk;\delta) & = b_0 (  - v_{_F}^2 r^2 + (\delta c_{\sharp} - \lambda_+)^2 ) r^2 \cos{2\theta}.
\end{aligned}
\end{equation*}
Then, the parametric curve:
\begin{equation*}
    \gamma_+(\theta)=\bigl(\Re a_+(\bk(r,\theta);\delta), \q\Im a_+(\bk(r,\theta);\delta) \bigr), \q \theta\in[0,2\pi],
\end{equation*}
is an ellipse revolving around the origin twice. Since 
\begin{equation*}
    \begin{aligned}
       & \biggl(  b_0 ( v_{_F}^2 r^2 + (\delta c_{\sharp} - \lambda_+)^2 ) r^2 \biggr) \biggl( b_0 (  - v_{_F}^2 r^2 + (\delta c_{\sharp} - \lambda_+)^2 ) r^2  \biggr) \\
       =& b_0^2r^4\biggl( - v_{_F}^4 r^4 + (\delta c_{\sharp} - \lambda_+)^4 \biggr) \to 0_-, \q r\to 0_+,
    \end{aligned}
\end{equation*}
the winding number of $a_+(\bk;\delta)$ is $-2$. This confirms $\Pi_3 =\Pi_4 = 1$ when $\sgn(\delta c_{\sharp}) >0$. The calculations in other cases are similar.

\subsubsection{Symmetries}
Besides, let us consider the grading operator $G:$
$$
	G = \begin{pmatrix}
		0 \q & 0\q  & 0 \q & 1 \\
	    0 \q & 0 \q & -1 \q & 0 \\
	    0 \q & 1 \q & 0 \q & 0\\
	    -1 \q & 0 \q & 0 \q & 0
	\end{pmatrix}
$$
satisfying $G^2 = -I$. The bifurcation matricies satisfy:
\begin{equation*}
	G^* \Tilde{B}_1^{\delta}(\bk) G = \overline{\Tilde{B}_1^{\delta}(\bk)}; \qq G^* \Tilde{B}_2^{\delta}(\bk) G = - \overline{\Tilde{B}_2^{\delta}(\bk)}.
\end{equation*}
Such difference between $\Tilde{B}_1^{\delta}(\bk)$ and $\Tilde{B}_2^{\delta}(\bk) $ indicates that the first order approximation is more ``symmetric" \cite{Bal2023B}. Therefore, the topology must also be considered with higher-order terms.

}

\subsection{Parities}\label{sec-topo}

The differences between perturbed bulk operators $\mch_{\pm}^{\delta}$ can be explained by the parities of eigenstates. Such parities originate in the $\frac{2}{3}\pi$ rotation symmetry and the $\mcp\mct$ symmetry.  The results regarding parities closely align with the findings on local charges discussed in the previous subsection that the upper pair bands and the lower pair bands are always the same within each pair and different from each other.

{
\begin{theorem}{\bf (Parities of eigenstates)}
    Let $\mch_V = -\Delta + V(\bx)$ be an operator possessing a fourfold degeneracy at the $\Gamma$ point as in Theorem \ref{thm-doublecone}. Assume that $W(\bx)$ is a folding symmetry breaking potential as above in (\ref{def-perturbation}) and satisfies the non-degeneracy condition (\ref{eqn-cSharp}):
    $$  c_{\sharp} = \langle \phi_1(\bx) , W(\bx)\phi_1(-\bx) \rangle_{L^2(\Omega)} \neq 0.$$
    The energy surfaces $\big\{\mcs_{n_*+j} \big\}_{j=1,2,3,4}$ of perturbed operator $\mch^{\delta} = -\Delta + V(\bx) + \delta W(\bx)$ will open a gap for $\delta$ sufficiently small as in Theorem \ref{thm-perturbation}.
    Then the parities of eigenfunctions corresponding to these four branches at $\bk={\bf 0}$ satisfy one of the following situations:
    \begin{enumerate}
        \item $\delta c_{\sharp}>0$: the upper two bands' eigenfunctions are odd and the lower two are even;
        \item $\delta c_{\sharp}<0$: the upper two bands' eigenfunctions are even and the lower two are odd.
    \end{enumerate}
\end{theorem}

In this subsection, we will prove this theorem by first solving for the leading terms in the corresponding expressions, then extending the results of these main terms to the exact eigenfunctions through symmetry arguments.
}

Solve $\bigl(\lambda(\bk;\delta) I +B^{\delta}_1(\bk)\bigr) \bm{P}(\bk;\delta) = 0$ and obtain $\lambda_{\pm} (\bk;
\delta)= \pm \sqrt{4\Vert \bk\cdot  \bm{v}_{\sharp}\Vert^2+\delta^2c_{\sharp}^2 }$. Both $\lambda_{\pm}(\bk;\delta)$ are of multiplicity two. The corresponding eigenvectors are:
\begin{equation*}
	\bm{P}_{\pm,1}(\bk;\delta) = \frac{1}{\sqrt{2}\lambda_{\pm}}\begin{pmatrix}
		  \lambda _{\pm} \vspace{0.1cm} \\ \overline{2i\bk\cdot  \bm{v}_{\sharp}}\\ -\delta c_{\sharp} \\ 0	 
	\end{pmatrix};\qq
	\bm{P}_{\pm,2}(\bk;\delta) = \frac{1}{\sqrt{2}\lambda_{\pm}}\begin{pmatrix}
		0 \\ -\delta c_{\sharp} \vspace{0.1cm} \\ -2i\bk\cdot  \bm{v}_{\sharp} \\ \lambda_{\pm}
	\end{pmatrix}.
\end{equation*}
When $\bk=0$, we obtain that:
\begin{equation*}
	\bm{P}_{\pm,1}({\bf 0};\delta) = \frac{1}{\sqrt{2}}\begin{pmatrix}
		 1 \vspace{0.1cm} \\ 0\\ \mp \sgn({\delta c_{\sharp}}) \\ 0	 
	\end{pmatrix};\qq
	\bm{P}_{\pm,2}({\bf 0};\delta) = \frac{1}{\sqrt{2}}\begin{pmatrix}
		0 \\ \mp\sgn(\delta c_{\sharp}) \vspace{0.1cm} \\ 0 \\ 1
	\end{pmatrix}.
\end{equation*}
This means the main terms of the corresponding eigenstates at $\bk = {\bf 0}$ are:
\begin{equation*}
	\begin{aligned}
		\phi_{\pm,1}(\bx;\delta) = \frac{1}{\sqrt{2}}\bigl(  \phi_1(\bx) \mp \sgn(\delta c_{\sharp})\phi_3(\bx) \bigr);\\
		\phi_{\pm,2}(\bx;\delta) = \frac{1}{\sqrt{2}}\bigl(  \phi_4(\bx) \mp \sgn(\delta c_{\sharp})\phi_2(\bx) \bigr) .
	\end{aligned}
\end{equation*}

The first interesting observation is that  $\phi_{\pm,2}(\bx) = \mct[\phi_{\pm,1}](\bx)$, which means that these two states are connected by time-reversal symmetry. 

Besides, note that $\phi_3(\bx) = \phi_1(-\bx)$ and $\phi_2(\bx) = \phi_4(-\bx)$. Thus, $\phi_{+,1}(\bx)$ and $\phi_{+,2}(\bx)$ are even when $\sgn(\delta c_{\sharp}) =-1$, and are odd when $\sgn(\delta c_{\sharp}) =1$. The parity of $\phi_{-,1}(\bx)$ and $\phi_{-,1}(\bx)$ are opposite to them. This means that the parities of the upper two bands at the $\Gamma$ point are always the same and opposite to those of the lower two bands. And the parities change when $\sgn(\delta c_{\sharp})$ changes.

These observations are true for not only the main terms but also the true eigenstates at the $\Gamma$ point when adding small folding symmetry breaking perturbations. This result originates in the $C_6$ symmetry and $\mcp\mct$ symmetry of the operator $\mch^{\delta}$ and the $C_6$ symmetry of the $\Gamma$ point:
\begin{equation}
	\chi = \bigoplus_{l=0}^5 \Tilde{\chi}_{l}; \qq \Tilde{\chi}_{l} = \big\{f \in \chi: \Tilde{\mcr}[f](\bx) = e^{\frac{\pi i}{3}l}f(\bx)\big\}.
\end{equation}
$\Tilde{\mcr}$ is the $\frac{\pi}{3}$-rotation operator: $\Tilde{\mcr}[f](\bx) = f(\Tilde{R}^*\bx)$, where
\begin{equation*}
	\Tilde{R}^* = \begin{pmatrix}
		\frac{1}{2} & \frac{\sqrt{3}}{2}\\
		-\frac{\sqrt{3}}{2} & \frac{1}{2}
	\end{pmatrix}
\end{equation*}
represents the anticlock $\frac{\pi}{3}$-rotation in $\R^2$. $\mch^{\delta}$ is commutative with $\Tilde{\mcr}$. Thus, each $\Tilde{\chi}_l$ is an invariant subspace of $\mch^{\delta}$. These characteristic subspaces of $\Tilde{\mcr}$ have the following properties associated with $\mcp$ and $\mct$ symmetries.
\begin{itemize}
	\item Each $\Tilde{\chi}_{l}$ is $\mcp$ invariant: because $\mcp = \Tilde{\mcr}\circ\Tilde{\mcr}\circ\Tilde{\mcr} $, any $f(\bx)$ in $\Tilde{\chi}_{l}$ is even when $j$ is even and odd when $j$ is odd.
	\item $\Tilde{\chi}_{l}$ and $\Tilde{\chi}_{6-l}$ are mapped to each other by $\mct$: if $f(\bx)$ is in $\Tilde{\chi}_l$, then $\mct[f](\bx)$ is in $\Tilde{\chi}_{6-l}$; specially if $f(\bx)$ is in $\Tilde{\chi}_0$, $\mct[f](\bx)$ is in $\Tilde{\chi}_0$.
\end{itemize}  

The eigenstates $\phi_l(\bx)$ of $H^0 = \mch_V$ are in characteristic subspaces of $\mcr$ and $\mcv_1$ as in the second conclusion in Theorem \ref{thm-doublecone}. After some linear combinations, they can be rearranged into:
\begin{equation*}
\begin{aligned}
	\Tilde{\phi}_1(\bx) = \frac{1}{\sqrt{2}}\bigl( \phi_1(\bx) +\phi_3(\bx) \bigr) \in \Tilde{\chi}_4;\\
	\Tilde{\phi}_2(\bx) = \frac{1}{\sqrt{2}}\bigl( \phi_4(\bx) +\phi_2(\bx) \bigr) \in \Tilde{\chi}_2;\\
	\Tilde{\phi}_3(\bx) = \frac{1}{\sqrt{2}}\bigl( \phi_1(\bx) -\phi_3(\bx) \bigr) \in \Tilde{\chi}_1;\\
	\Tilde{\phi}_4(\bx) = \frac{1}{\sqrt{2}}\bigl( \phi_4(\bx) -\phi_2(\bx) \bigr) \in \Tilde{\chi}_5.
\end{aligned}	
\end{equation*}

By the continuity of the spectrum concerning $\delta$ \cite{Fefferman2014}, the four branches $\big\{ E_{b+j}(\bk;\delta)$ $: \bk \in \Omega^*\big\}_{j=1,2,3,4}$ possess eigenstates in $\Tilde{\chi}_1$, $\Tilde{\chi}_2$, $\Tilde{\chi}_4$, $\Tilde{\chi}_5$ respectively and decompose into a pair of twofold degeneracy since $\Tilde{\chi}_l$ and $\Tilde{\chi}_{6-l}$ are connected by $\mct$ symmetry. This together with the situation of main terms of the eigenstates confirms the theorem.

\section{A typical numerical example}\label{sec-num}

A typical example in physics is the kind of $\mcp\mct$ symmetric structure proposed by Wu and Hu \cite{Hu2015}. We numerically study the associated edge Schr\"odinger operator:
\begin{equation*}
	\mch_{edge} = -\Delta + W_{edge}(\bx).
\end{equation*} 
Here the potential $W_{edge}(\bx)$ is a piecewise constant function in $L^2(\R^2)$. In the analysis in the above sections, we always suppose all the potentials are smooth. For general discontinuous potentials in $L^2(\R^2)$, Theorem \ref{thm-perturbation}, Theorem \ref{thm-second-order-cone}, and the multiscale expansions are still valid, but other results should be treated carefully after smoothing, especially for the error estimation.

The parameters related to the honeycomb lattice are:
\begin{equation*}
	\buu_1 = \begin{pmatrix}
		\frac{\sqrt{3}}{2} \\ 
		\frac{1}{2}
	\end{pmatrix}, \q
	\buu_2 = \begin{pmatrix}
		\frac{\sqrt{3}}{2} \\ 
		-\frac{1}{2}
	\end{pmatrix}, \q 
	\buu_3 = \buu_2-\buu_1 = \begin{pmatrix}
		0 \\ -1
	\end{pmatrix};
\end{equation*}
\begin{equation*}
	\bk_1 = \frac{4\pi}{\sqrt{3}}\begin{pmatrix}
		\frac{1}{2} \\ 
		\frac{\sqrt{3}}{2}
	\end{pmatrix}, \q
	\bk_2 = \frac{4\pi}{\sqrt{3}}\begin{pmatrix}
		\frac{1}{2} \\ 
		-\frac{\sqrt{3}}{2}
	\end{pmatrix}.
\end{equation*}
The edge direction is $\bm{l}_2 = \bk_2.$
The potential is
\begin{equation*}
	W_{edge}(\bx) = \left\{ 
	\begin{aligned}
		&g(\bx;\frac{1.1}{3}),\q \bx\cdot\bm{l}_2 \geq 0; \\
		&g(\bx;\frac{0.9}{3}),\q else.
	\end{aligned}\right.
\end{equation*}
Here $g(\bx,\frac{1.1}{3})$ and $g(\bx,\frac{0.9}{3})$ are the perturbed bulk potentials on the two sides of the edge with different topologies. They are deformed from $g(\bx;\frac{1}{3})$ -- a potential possessing all the properties of the super honeycomb lattice potential except for the smoothness. They are shrunk and expanded super honeycomb lattice potentials respectively, as shown in Figure \ref{figure-energy-curve}. They can be constructed by rotation of dimers as below:
\begin{equation*}
	g(\bx;r) = f(\bx;r) + \mcr f(\bx;r) + \mcr^2 f(\bx;r).
\end{equation*}
The potential of a group of dimers is:
\begin{equation*}
	f(\bx;r) = a(\bx - \frac{1}{2}r\buu_3) + a(\bx + \frac{1}{2}r\buu_3)
\end{equation*}
with $a(\bx)$ $\buu_1$ and $\buu_2$ doubly periodic. In the unit cell, it is
\begin{equation*}
	a(\bx)= \left\{ 
	\begin{aligned}
		& 10 , \qq     |x-\frac{1}{2}(\buu_1 + \buu_2)|<0.1,\\
		& 300, \qq  else.
	\end{aligned}\right.
\end{equation*}
 
 We use the finite element method \cite{Guo2022} to get the picture of $\mch_{edge}$'s spectrum and eigenstates as shown in Figure \ref{figure-energy-curve} and Figure \ref{figure-ev}. (b) of Figure \ref{figure-energy-curve} shows the existence of two gapped edge states. (a) and (b) of  Figure \ref{figure-ev} are corresponding eigenstates. (c)-(f) of Figure \ref{figure-ev} are a basis of eigenstates of the operator
 \begin{equation*}
 	\mch_{bulk} = -\Delta + g(\bx;\frac{1}{3})
 \end{equation*}
 with energy $E_D$ marked in (b) of Figure \ref{figure-energy-curve}. The operator $\mch_{bulk}$ has a fourfold degeneracy on its energy band at the $\Gamma$ point as in  Theorem \ref{thm-doublecone} with eigenstates (c)-(f). The edge modes in (a) and (b) are modulations of (c) and (d) in  Figure \ref{figure-ev}.

\section*{Acknowledgments}
We would like to acknowledge Guochuan Thiang and Borui Miao for inspiring discussions and precious suggestions and Shuo Yang for providing help in establishing numerical simulations.

\appendix
\section{The form of $b(\bk)$} \label{appen}

From the expression of $b(\bk)$ in (\ref{eqn-b1}), we know that:
\begin{equation*}
	\begin{aligned}
	    b(\bk) & = \langle   \phi_1(\bx), 2i\bk \cdot \nabla(\mch_V-E_{_D})^{-1} \mcq_{\perp}  2i\bk \cdot \nabla   \phi_2(\bx) \rangle _{L^2(\Omega)}\\
		& = \langle 2i\bk \cdot \nabla  \phi_1(\bx), (\mch_V-E_{_D})^{-1} \mcq_{\perp}  2i\bk \cdot \nabla   \phi_2(\bx) \rangle _{L^2(\Omega)} \\
		& = \langle\mcr\biggl( 2i\bk \cdot \nabla  \phi_1(\bx)\biggr), \mcr\biggl((\mch_V-E_{_D})^{-1} \mcq_{\perp}  2i\bk \cdot \nabla   \phi_2(\bx)\biggr) \rangle _{L^2(\Omega)} \\
		& = \langle 2iR^*\bk \cdot \nabla  \mcr\phi_1(\bx), (\mch_V-E_{_D})^{-1} \mcq_{\perp}  2iR^*\bk \cdot \nabla \mcr  \phi_2(\bx) \rangle _{L^2(\Omega)}\\
		& = \tau^2 \langle 2iR^*\bk \cdot \nabla  \phi_1(\bx), (\mch_V-E_{_D})^{-1} \mcq_{\perp}  2iR^*\bk \cdot \nabla   \phi_2(\bx) \rangle _{L^2(\Omega)}\\
		& = \bar{\tau} b(R^*\bk),
	\end{aligned}
\end{equation*}
which means it satisfies:
\begin{equation}\label{eqn-appen}
	b(R^*\bk) = \tau b(\bk).
\end{equation}
Here 
$$R^*= \begin{pmatrix}  -\frac{1}{2} & -\frac{\sqrt{3}}{2}  \\  \frac{\sqrt{3}}{2} & -\frac{1}{2}  \end{pmatrix}.$$

We can write $b(\bk)$ into a polynomial of $b(\bk) = \bk^T \begin{pmatrix} A & C\\ 0&B
	\end{pmatrix}\bk
$.
We can obtain:
\begin{equation*}
	b(R^*\bk)  = \bk^{^{T}} R \begin{pmatrix} A & C\\ 0&B
	\end{pmatrix} R^* \bk.
\end{equation*}
Combined with (\ref{eqn-appen}):
\begin{equation}
	\begin{pmatrix}
		\frac{1}{4} & \frac{3}{4} & -\frac{\sqrt{3}}{4} \\
		\frac{3}{4} & \frac{1}{4} & \frac{\sqrt{3}}{4}\\
		\frac{\sqrt{3}}{2} & -\frac{\sqrt{3}}{2} & -\frac{1}{2}
	\end{pmatrix}\begin{pmatrix}
		A \\ B\\ C
	\end{pmatrix} = \tau \begin{pmatrix}
		A \\ B\\ C
	\end{pmatrix}.
\end{equation}
The solution is
\begin{equation}
	\begin{pmatrix}
		A \\ B\\ C
	\end{pmatrix} = b_0\begin{pmatrix}
		i \\ -i\\ 2
	\end{pmatrix}.
\end{equation}
Therefore, we can write $b(\bk) = b_0 \bk^T \begin{pmatrix}
    i & 2 \\ 0 & -i
\end{pmatrix} \bk$, with $b_0$ a nonzero constant.



\end{document}